\documentclass[11pt,a4paper]{article}
\usepackage{jcappub}

\usepackage[colorlinks = true,
            linkcolor = blue,
            urlcolor  = blue,
            citecolor = blue,
            anchorcolor = blue]{hyperref}
\usepackage{amsmath,amssymb}
\usepackage[capitalize]{cleveref}
\usepackage{graphicx}
\usepackage{slashed}
\usepackage{enumitem}
\usepackage[dvipsnames]{xcolor}

\newcommand{\dd}{\mathrm{d}}

\newcommand{\keV}{\,\mathrm{keV}}

\newcommand{\GeV}{\,\mathrm{GeV}}

\newcommand{\agree}[1]{{\color{black} #1}}

\begin{document}

\title{Structure Formation Limits on Axion-Like Dark Matter}

\author[a]{Sven Baumholzer}
\author[b,c]{Vedran Brdar}
\author[a]{Enrico Morgante}

\affiliation[a]{PRISMA Cluster of Excellence \& Mainz Institute for Theoretical Physics,
	Johannes Gutenberg University, Staudingerweg 7, 55099 Mainz, Germany}
\affiliation[b]{
	Fermi National Accelerator Laboratory, Batavia, IL, 60510, USA}
\affiliation[c]{Northwestern University, Department of Physics \& Astronomy, 2145 Sheridan Road, Evanston, IL 60208, USA}

\emailAdd{baumholz@uni-mainz.de}
\emailAdd{vedran.brdar@northwestern.edu}
\emailAdd{emorgant@uni-mainz.de}


\abstract{
We derive structure formation limits on dark matter (DM) composed of keV-scale axion-like particles (ALPs), produced via freeze-in through the interactions with photons and Standard Model (SM) fermions.
We employ Lyman-alpha (Ly-$\alpha$) forest data sets as well as the observed number of Milky Way (MW) subhalos. We compare results obtained using Maxwell-Boltzmann and quantum statistics for describing the SM bath. It should be emphasized that the presence of logarithmic divergences complicates the calculation of the production rate, which can not be parameterized with a simple power law behaviour.
The obtained results, in combination with X-ray bounds, exclude the possibility for a photophilic ``frozen-in" ALP DM with mass below $\sim 19\keV$. For the photophobic ALP scenario, in which DM couples primarily to SM fermions, the ALP DM distribution function is peaked at somewhat lower momentum and hence for such realization we find weaker limits on DM mass. Future facilities, such as the upcoming Vera C. Rubin observatory, will provide measurements with which the current bounds can be significantly improved to $\sim 80\keV$.}

\begin{flushright}
	MITP-20-073\\
	nuhep-th/20-14
\end{flushright}

\maketitle


\section{Introduction}
The waning of the WIMP \cite{Arcadi:2017kky} is redirecting (astro)particle DM research towards alternative 
DM realizations. For instance, non-thermally produced light DM in form of sterile neutrinos \cite{Dodelson:1993je,Shi:1998km,Merle:2013wta,Brdar:2017wgy,deGouvea:2019phk}, fuzzy DM \cite{Lee:2017qve,Hui:2016ltb}, hidden photons \cite{Redondo:2008ec,Hambye:2019dwd} and ALPs \cite{Arias:2012az,Jaeckel:2014qea,Im:2019iwd} is not only receiving a significant attention in light of existing  and forthcoming terrestrial experiments \cite{Ouellet:2018beu}, but also has the potential to resolve discrepancies between DM observations and simulations \cite{Marsh:2013ywa}. Additionally, non-thermal DM candidates can be related to new solutions of the old particle physics problems, such as the hierarchy problem in the case of relaxion DM~\cite{Fonseca:2018kqf, Fonseca:2020pjs, Banerjee:2018xmn}.

 In this work, we focus on ALP DM with $\mathcal{O}(1-100)\keV$ mass, which is the scale motivated by hints in X-ray data \cite{Bulbul:2014sua,Boyarsky:2014jta} as well as the recent  measurement of an excess in electron recoil spectrum performed by the XENON1T collaboration \cite{Aprile:2020tmw}.
In fact, the former measurement can be explained by ALP coupling to photons \cite{Jaeckel:2014qea}, while for the latter, ALP coupling to fermions (namely electrons) suffices \cite{Fonseca:2020pjs}. In light of these statements, we study  ALP production via freeze-in through feeble interactions with photons and SM fermions. Our primary mission is to compute the structure formation limits that have not been derived to date for keV-scale ALPs; for this purpose we employ recent Ly-$\alpha$ data as well as the observed number of MW subhalos.
Such structure formation limits are widely scrutinized for light sterile neutrino DM \cite{Merle:2015vzu, Schneider:2016uqi}. While using the results from these studies would allow one to get a very rough estimate on the structure formation limits for ALP DM, we find it valuable, especially in light of aforementioned experiments that may start to detect DM, to perform a dedicated study and determine with large precision a viable parameter space for the considered model. We also point out that for bosonic DM such bounds are important  irrespective of their strength; in contrast, structure formation limits on fermionic DM, namely sterile neutrinos, could be overridden by those stemming from phase space, \emph{i.e.} Pauli blocking \cite{Tremaine:1979we}. 
 
 The paper is organized as follows. In \cref{sec:model} we introduce the model and discuss channels for ALP DM production. We then detail computations for the approaches in which Maxwell-Boltzmann (\ref{subsec:MB}) and quantum (\ref{subsec:quantum}) statistics are employed to describe particles in the SM bath. Then, in \cref{sec:data} we discuss experimental observations that allow us to constrain the ALP DM parameter space. In \cref{sec:discussion} we discuss theoretical aspects of ALP DM in the early Universe, focusing on the interplay between misalignment and freeze-in production. In \cref{sec:results} our main results are presented. Finally, we summarize in \cref{sec:conclude}.

\section{The Model and DM production.}
\label{sec:model}

The relevant part of the Lagrangian reads
\begin{align}
\mathcal{L}  =\, & \frac{1}{2}\partial_\mu a \, \partial^\mu a +\frac{1}{2}m_a^2 a^2 + \bar{f} (i\slashed{\partial} - m_f) f - \frac{1}{4}F_{\mu\nu} F^{\mu\nu} \nonumber \\ & - q \, e \, \bar{f} \slashed{A} \, f  
 + \frac{c_{a\gamma\gamma}}{4 f_a} a F_{\mu\nu} \widetilde F^{\mu\nu} +
\frac{c_{aff}}{f_a} \partial_\mu a \,\bar{f} \gamma^\mu\gamma_5 f \,.
\label{eq:lag}
\end{align}
Here, we first write down the kinetic and mass term for the pseudoscalar ALP field $a$; the following three terms describe quantum electrodynamics, namely a SM fermion $f$ with an electric charge $q$ interacting with photon field $A^\mu$ whose corresponding (dual) field strength tensor is denoted by $F^{\mu\nu}$ ($\tilde{F}^{\mu\nu}$); the last two terms describe the interaction of ALPs with $A^\mu$ and $f$, respectively.
We will chiefly analyze ALP coupling to photons and fermions separately, \emph{i.e.} working under the assumption that either $c_{aff}$ or $c_{a\gamma\gamma}$ vanishes. These scenarios are dubbed as \emph{photophilic} and \emph{photophobic}, respectively.

Regarding the photophilic case, at temperatures above $\sim 160\GeV$ the EW symmetry is restored, and one should work in the $\{B_\mu, W^i_\mu\}$ basis. We assume for simplicity that ALP couples only to the $U(1)$ gauge fields through
\begin{equation}\label{eq:coupling to B}
\frac{c_{aBB}}{4f_a} a B_{\mu\nu} \widetilde B^{\mu\nu} \,.
\end{equation}
The inclusion of the $W_\mu^i$ and gluon fields would increase the production rate of ALPs, but we chose to limit ourselves to the coupling to photons, which is the scenario typically considered
in the literature for heavy ALP searches \cite{Feng:2018pew,Dolan:2017osp,Alekhin:2015byh,Brdar:2020dpr}.\\

In the photophobic case, using the equations of motion, one can replace the last term in \cref{eq:lag} with the term proportional to the fermion mass \cite{Bauer:2017ris}, $(2m_f/f_a)\, a \,\bar{f} \gamma_5 f$. This does not imply that such coupling is only operative once electroweak symmetry is broken, for $T\lesssim 160\GeV$. On the contrary, after a phase redefinition of the quark and Higgs fields, a term $i y_t (c_t a/f_a) \bar Q_3 H t_R$ is generated~\cite{Salvio:2013iaa}, where $Q_3$ is the third-generation left-handed doublet, $t_R$ is the right handed top quark, and $c_t$ is a free dimensionless parameter. Such a term, thanks to the large top Yukawa, contributes very efficiently to the ALP production, and it can be the dominant source when all ALP couplings to SM particles have the same order of magnitude.
In this work, however, we limit ourselves to the sub-TeV values of the reheating temperature in the photophobic scenario, namely $T_\mathrm{RH}<160\GeV$. In that regime, the top quark is Boltzmann suppressed and does not contribute to ALP production. Due to the $m_f$ suppression of the ALP-fermion coupling, the relevant fermions for ALP production are bottom and charm quarks as well as tau leptons.\\

\begin{figure}
\centering
\includegraphics[scale=1]{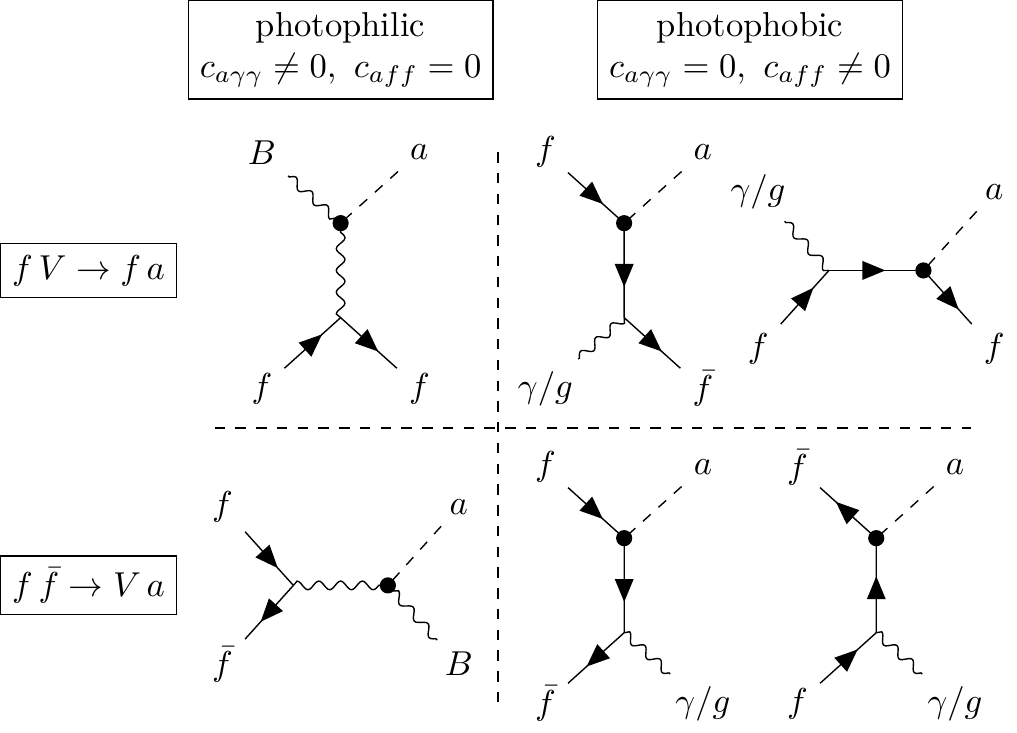}
\caption{Tree level Feynman diagrams for the two processes of interest, $f V\to f a$ (top) and $\bar{f}f\to V a$ (bottom), in the photophilic (left) and photophobic (right) cases. }
\label{fig:feynman}
\end{figure}

The kinetic and interaction terms in \cref{eq:lag} facilitate ALP production from thermalized fermions and vector bosons $V$ (the photon, a gluon, or the $U(1)_Y$ gauge field) through $V f \to a f$ and $\bar{f} f \to V a$ channels. For both of these processes there are one photon- and two $f$-mediated Feynman diagrams contributing at the tree-level%
\footnote{We have also considered ALP production from Higgs bosons that couple to $U(1)_Y$ gauge field. In the photophilic case we found such contribution to be negligible, namely at the percent level with respect to the production through diagrams shown in \cref{fig:feynman} (upon summing over all fermion species). Note that in the photophobic scenario, given the low $T_\text{RH}$ under our consideration, the Higgs boson abundance in the bath is Boltzmann suppressed.}(see Fig.~\ref{fig:feynman}).
We note that for intermediate scenarios where $c_{aff}\sim c_{a\gamma\gamma}$, our results turn out to be driven by couplings to photons, namely they match the photophilic scenario. This is because, in the photophobic case, the ALP interaction operator is proportional to the fermion mass, and thus the production rate is suppressed  with respect to the photophilic case by a factor $m_f/T_\mathrm{RH}$.

Independently of the scenario under consideration, the production of keV-scale ALP DM through the processes depicted in \cref{fig:feynman} should occur  via ``freeze-in" \cite{Hall:2009bx}, because for $m_a\gtrsim 100\,\mathrm{eV}$ the freeze-out would lead to DM relic abundance that greatly exceeds present measurements~\cite{Jaeckel:2014qea}. The absence of ALP DM thermalization simply implies that the reheating temperature, $T_{\text{RH}}$, should not exceed the temperature at which ALP decouples from the thermal plasma. 
The decoupling temperature can be estimated by calculating the scattering rate, $\Gamma$, from the momentum integrated collision term given in \cref{eq:Boltzmann_equation} and the number density of SM fermions, $n_f = 72\zeta(3)T^3/\pi^2$. In the photophilic case it yields
	\begin{align}\label{eq:interaction rate photophilic}
\Gamma = \frac{\int \dd^3 p \, C(x,r)}{n_f} = \frac{5 g'^2 T^3}{864 \pi^3 \zeta(3) \cos^2\theta_W} \frac{c_{a\gamma\gamma}^2}{f_a^2} \left( 23 -24\gamma_E -12 \log\left(\frac{11 g'^2}{48}\right) \right)\,,
	\end{align}
where $\theta_W$ ($\cos\theta_W \sim 0.88$) is the weak mixing angle and $\gamma_E$ is the
Euler-Mascheroni constant. For the numerical values of QED and $U(1)_Y$ coupling constants, $e$ and $g'$, we use $\sqrt{4\pi/137}$
and $0.35$, respectively. We neglect the running of these parameters.

Fixing the coupling $c_{a\gamma\gamma}/f_a$ (using the distribution of \cref{eq:f photophilic Teresi} and the expression for $\Gamma$ in above equation) in order to match the observed relic abundance of $\Omega_{\text{DM}} h^2\approx 0.12$ we obtain
\begin{align}\label{eq:TRHversus}
\frac{T_\text{RH}}{T_\text{dec}} \simeq 2.3\times 10^{-4} \left(\frac{g_*(T)}{106.75}\right) \left(\frac{10\keV}{m_a}\right).
\end{align}
Hence, for $m_a\sim \mathcal{O}$(keV) it is guaranteed that the ALP DM never thermalizes. 

For the photophobic scenario we note that $\Gamma / H$ is largest for temperatures $T\simeq m_f$. However, as we will show, demanding that $\Omega_\text{DM}h^2$
is matched, one requires $c_{aff}/f_a \lesssim 10^{-9}\,\GeV^{-1}$ for $m_a = \mathcal{O}(10\keV)$ and with such feeble couplings it is guaranteed that photophobic ALP DM does not thermalize with SM species.

Using the distribution functions presented later in the text, for the contribution from $\bar{f}f\to B a$ in photophilic scenario we find 
\begin{align} 
 \Omega_\text{DM}^{\bar{f}f\to B a} h^2 \approx  0.12 \left(\frac{106.75}{g_*}\right)^{3/2} \left( \frac{c_{a\gamma\gamma}/f_a}{10^{-17}\GeV^{-1}} \right)^2 
			\left( \frac{m_\text{DM}}{10 \,\text{keV}} \right) \left( \frac{T_\text{RH}}{6.7\cdot10^{16}\, \text{GeV}} \right) \,,
\label{eq:omega-h2-1}
\end{align}
and the $f B \to f a$ process yields
\begin{align}
\Omega_\text{DM}^{f B \to f a} h^2  
		       \approx \, 0.12  \left(\frac{106.75}{g_*}\right)^{3/2} \left( \frac{c_{a\gamma\gamma}/f_a}{10^{-17}\GeV^{-1}} \right)^2 \left( \frac{m_\text{DM}}{10\, \text{keV}} \right)  \left( \frac{T_\text{RH}}{2.7 \cdot10^{15}\, \text{GeV}}\,  \right) \,.
\label{eq:omega-h2-2}
\end{align}
Including both contributions we get
\begin{align}\label{eq:omega-h2-3}
\Omega_\text{DM} h^2  
		       \approx \, 0.12  \left(\frac{106.75}{g_*}\right)^{3/2} \left( \frac{c_{a\gamma\gamma}/f_a}{10^{-17}\GeV^{-1}} \right)^2 \left( \frac{m_\text{DM}}{10\, \text{keV}} \right)  \left( \frac{T_\text{RH}}{2.6 \cdot10^{15}\, \text{GeV}}\,  \right) \,.
\end{align}

Clearly, the scattering process $f B \to f a$ represents the dominant production channel in the photophilic scenario and hence will also give the dominant impact in the determination of structure formation limits. The reason for that lies in the 
logarithmic enhancement arising due to regularization of t-channel $B$-mediated process with a thermal gauge boson mass.
 
The equivalent equations to \cref{eq:omega-h2-1,eq:omega-h2-2} for the photophobic scenario read
\begin{equation}\label{eq:omega-h2-1-photophobic}
\Omega_\text{DM}^{\bar{f}f\to V a} h^2 \approx
0.12 \left(\frac{80}{g_*}\right)^{3/2} \left( \frac{m_\text{DM}}{10\, \text{keV}} \right)  \left(\frac{\sum_f\kappa_f}{38.9\GeV} \right) \left( \frac{c_{aff}/f_a}{9.6\cdot10^{-11}\GeV^{-1}} \right)^2\,,
\end{equation}
and 
\begin{equation}\label{eq:omega-h2-2-photophobic}
\Omega_\text{DM}^{f V \to f a} h^2 \approx
0.12 \left(\frac{80}{g_*}\right)^{3/2} \left( \frac{m_\text{DM}}{10\, \text{keV}} \right)  \left(\frac{\sum_f\kappa_f}{38.9\GeV} \right) \left( \frac{c_{aff}/f_a}{1.2\cdot10^{-10}\GeV^{-1}} \right)^2\,,
\end{equation}
while the sum gives
\begin{equation}\label{eq:omega-h2-3-photophobic}
\Omega_\text{DM} h^2 \approx
0.12  \left(\frac{80}{g_*}\right)^{3/2} \left( \frac{m_\text{DM}}{10\, \text{keV}} \right)  \left(\frac{\sum_f\kappa_f}{38.9\GeV} \right) \left( \frac{c_{aff}/f_a}{7.6\cdot10^{-11}\GeV^{-1}} \right)^2\,.
\end{equation}
The factor $\kappa_f$ is equal to $m_f \, n_c \, q^2 e^2$ for leptons and to $m_f \, n_c \, q^2 e^2 + 4\, m_f \, g_s^2$ for quarks, where $n_c$ is the number of colors of the fermion $f$ and $q$ its electric charge. For quarks, we took into account the diagrams with an external gluon. In this case, $g_s$ is the strong gauge coupling which we fixed to  $1.31$ (value obtained by averaging $\alpha_s$ over the relevant energy range between the b-quark and $Z$ mass), and the factor $4$ comes from the trace over the $SU(3)$ generators. The parameter $\kappa_f$ takes the approximate values $\{0.16, 8.97, 29.0, 38.1, 38.9\}\GeV$ for tau, charm, bottom, the sum over these three, and the sum over all SM fermions (excluding the top), respectively.

One can infer from \cref{eq:omega-h2-1-photophobic,eq:omega-h2-2-photophobic} that both $\bar{f}f\to V a$ and $fV \to f a$ are of similar strength in the photophobic scenario.

The key ingredient to address structure formation is the ALP DM distribution function. It is obtained by solving the Boltzmann equation, which, for the process $1+2\to 3+a$, reads
\begin{align}
\left[\frac{\partial}{\partial t} - H p_a \frac{\partial}{\partial p_a} \right] f(p_a,t) = \mathcal{C}(p_a)\,,
\end{align}
with
\begin{align}\label{eq:collision term}
\mathcal{C}(p_a) = \frac{1}{2E}\int & \frac{d^3 p_1}{(2\pi)^32E_1} \frac{d^3 p_2}{(2\pi)^32E_2} \frac{d^3 p_3}{(2\pi)^32E_3} (2\pi)^4 \delta^4(P_1+P_2-P_3-P_a) \times \nonumber \\
& \times |\mathcal{M}|^2 f_1(E_1,T) f_2(E_2,T) (1-f_3(E_3,T))\,.
\end{align}
Here, $|\mathcal{M}|^2$ is the squared amplitude for the considered process.
In what follows, we present the computation of the ALP DM distribution function both for the case of Maxwell-Boltzmann and quantum statistics describing the SM bath.

\subsection{ALP  distribution function computed using Maxwell-Boltzmann statistics}\label{sec:MB stat}
\label{subsec:MB}
%
Using the Maxwell-Boltzmann statistics and assuming $f_i\ll 1$ for all particles involved, simplifies the Boltzmann equation considerably~\cite{Heeck:2017xbu}. Indeed, one can approximate $f_1(E_1) f_2(E_2) (1-f_3(E_3))\approx \exp[(E_1 + E_2)/T] = \exp(P_0/T)$, where $P = P_1 + P_2 = P_3 + P_4$ and $P_0$ is the first component (energy) of the 4-vector $P$. The collision term can then be factorized as
\begin{align}\label{eq:Boltzmann MB}
\mathcal{C}(p_a) & = \frac{1}{2E}\int \frac{d^4 P}{(2\pi)^4}\frac{e^{-P_0/T}}{2E_3} (2\pi) \, \delta(E_3 + E_a - P_0) \times \nonumber\\
& \times \int \frac{d^3 p_2}{(2\pi)^32E_2} \frac{d^3 p_3}{(2\pi)^32E_3} (2\pi)^4 \, \delta^4(P_1+P_2-P) \, |\mathcal{M}|^2\,.
\end{align}

Neglecting any CP violation, the second line is nothing but the reduced cross section, $\hat\sigma$, of the inverse process $3+a\to1+2$, multiplied by a phase space factor $\lambda(s,m_1^2,m_2^2)^{1/2}/s$. It is related to the usual cross section through 
$\hat{\sigma}=2 \, [\lambda(s,m_1^2,m_2^2)/s] \, \sigma$, where $s$ is a Mandelstam variable and $\lambda$ is the K{\"a}ll\'{e}n function. This function approximately reads $\lambda\approx s^2$ when masses $m_a,m_f \lesssim \sqrt{s}\sim T$ which is the case for the relevant epoch in the early Universe associated to ALP DM production. The second line of Eq.~\ref{eq:Boltzmann MB} is invariant under longitudinal boosts, which makes it simple to compute it in the center of momentum frame where $\vec p =0$.
We computed $\hat\sigma$ analytically at tree level considering the diagrams shown in Fig.~\ref{fig:feynman}. For QED, with one fermion of charge $\pm 1$, we obtained 
\begin{align}
&\hat\sigma_{\bar{f}f\to \gamma a}  =
\frac{e^2}{12\pi} \frac{c_{a\gamma\gamma}^2}{f_a^2} s
\left(1 + 2\frac{m_f^2}{s}\right) \sqrt{1-\frac{4m_f^2}{s}} \nonumber \\
& - \frac{2}{\pi} e^2 m_f^2 \left(2 \frac{c_{aff}^2}{f_a^2} + \frac{c_{aff}c_{a\gamma\gamma}}{f_a^2} \right)  \log\left(\frac{s - 2 m_f^2 - s \sqrt{1-4 m_f^2/s }}{2 m_f^2}\right)\,,
\label{eq:xsec1}
\end{align}
and
\begin{align}
&\hat\sigma_{f \gamma\to f a} = 
\frac{e^2}{16 \pi} \frac{c_{a\gamma\gamma}^2}{f_a^2}
s \left(1-\frac{m_f^2}{s}\right)^2  \left[4 \log \left(\frac{(s - m_f^2)^2}{s \, m_\gamma^2}\right) - 3 - 2 \frac{m_f^2}{s} + \frac{m_f^4}{s^2}\right] \nonumber \\
&  - \frac{e^2}{2\pi} \left( 2 \frac{c_{aff}^2}{f_a^2} + \frac{c_{aff}c_{a\gamma\gamma}}{f_a^2} \right)m_f^2  \left(1-\frac{m_f^2}{s}\right)
\left(2 \log \left(\frac{s}{m_f^2}\right) - 3 + 4 \frac{m_f^2}{s}  - \frac{m_f^4}{s^2}\right)\,.
\label{eq:xsec2}
\end{align}

Here, $m_\gamma\approx eT/3$ is the plasmon mass that serves as a regulator for the diagram including t-channel photon exchange. The expressions in \cref{eq:xsec1,eq:xsec2} are obtained for general cases of both $c_{a\gamma\gamma}$ and $c_{aff} \neq 0$ and have well defined limits if any of these two couplings vanishes. We note that \cref{eq:xsec2} holds for both 
$f$ and $\bar{f}$ scattering off photons, and thus it must be summed over twice.

The expressions in \cref{eq:xsec1,eq:xsec2} must be corrected to account for the presence of SM fermions. In the photophilic scenario, in which the ALP population is generated at $T\gg 160\GeV$, $e^2 c_{a\gamma\gamma}$ must be replaced by $g'^2 c_{aBB} = g'^2 c_{a\gamma\gamma}/\cos^2\theta_W$, and the result must be multiplied by the sum over SM fermions $g_Y\equiv \sum_f Y^2 n_c = 10$. The plasmon mass in this case is $(11/12)^{1/2} g' T$~\cite{Salvio:2013iaa}.
In the photophobic scenario, instead,
the electric charge $e^2$ should be multiplied by $q^2 n_c $ for leptons, and replaced with $q^2 n_c e^2 + 4 g_s^2$ for quarks, and finally summed over the SM fermions involved (typically $b, c, \tau$).

It is convenient to rewrite the Boltzmann equation~(\ref{eq:Boltzmann MB}) by defining dimensionless quantities $r=m_H/T$ and $x=p/T$, where $m_H$ is a reference mass which we fix to be the Higgs mass; its actual value is irrelevant as $m_H$ cancels in the calculation of physical quantities. With this redefinition one obtains the master formula in Eq.~(11) of \cite{Heeck:2017xbu} for ALP DM distribution function
\begin{align}
f(x,r)&=\frac{M_0}{16\pi^2\, m_H\,x^2} \int_{r_i}^{r_f}\,dr \int_{y^*}^\infty dy\, \hat{\sigma}\left(\frac{m_H^2\,y}{r^2}\right)\,  \text{Exp}\left[-x-\frac{y}{4x}\right]\,,
\label{eq:master_eq_momentum_distr}
\end{align}
where $M_0=M_\mathrm{Pl} \sqrt{45/(4\pi^3 g_*)}$ and $M_\mathrm{Pl}=1/G_N\approx 1.22\times 10^{19}\GeV$ is the Planck mass; $r_{i,f}$ are the lower and upper integration boundaries $m_H/T_\mathrm{RH}$ and $m_H/m_f$, respectively; $y^*= 4 r m_f^2/m_H^2$ for the fermion annihilation process and $y^*= r m_f^2/m_H^2$ for fermion scattering. Using this approach, the change of the number of relativistic degrees of freedom, $g_*$, is not taken into account but in the relevant temperature range (that is above QCD phase transition) such approximation is justified. 

In \cref{eq:master_eq_momentum_distr} and in the following, we stop at the first non-zero order in $m_f$, \textit{i.e.} we set $m_f=0$ everywhere apart from the cross section where we keep the factor $m_f^2$ multiplying the $c_{aff}^2$ terms. We also keep only the first non-zero $m_f$-dependent term inside the logarithms (see \cref{eq:xsec1,eq:xsec2}). The same holds for the integration limits $r_f$ and $y^*$, which can be sent, respectively, to infinity and to zero in the photophilic ALP case. Finally, we drop the interference terms $\propto c_{a\gamma\gamma} c_{aff}$, as we are going to consider separately the cases $c_{a\gamma\gamma}= 0$ and $c_{aff} = 0$. Note that the interference terms in \cref{eq:xsec1,eq:xsec2} are proportional to the corresponding $c_{aff}^2$ terms, and hence, even if considered, they would not lead to any major effect, in particular the energy dependence of the cross section would be unaltered.

The distribution function can thus be integrated analytically. We obtain
\begin{equation}\label{eq:f photophilic Teresi}
f(x) = \frac{g'^2}{12\pi^3 \cos^2\theta_W} g_Y M_0 T_\mathrm{RH}\frac{c_{a\gamma\gamma}^2}{f_a^2}  e^{-x} \left\{1 + \frac{3}{2} \left[1-4\gamma_E + 4 \log\left(\frac{48x}{11 g'^2}\right) \right] \right\}\,,
\end{equation}
in the photophilic  and
\begin{align}\label{eq:f photophobic Teresi}
f(x) & = \sum_f \frac{1}{2\pi^3}M_0 \, \kappa_f  \, \frac{c_{aff}^2}{f_a^2} \frac{e^{-x}}{x} \bigg\{
\left[ 2\sqrt{\pi x} \, (1+\log 2) \, \mathrm{erf}\left(\frac{1}{\sqrt{x}}\right) + 2\Gamma\left(0,\frac{1}{x}\right) \right]\nonumber \\
& + \left[\sqrt{\pi x}  \, \mathrm{erf}\left(\frac{1}{2\sqrt{x}}\right) + 2\Gamma\left(0,\frac{1}{4x}\right) \right] \bigg\} \,,
\end{align}
in the photophobic scenario. In these expressions, \emph{erf} is the error and $\Gamma(a,x)$ is the incomplete Gamma function.
In both (\ref{eq:f photophilic Teresi}) and~(\ref{eq:f photophobic Teresi}), the first term in the curly brackets comes from the fermion annihilation process $f\bar{f}\to V a$, while the second arises from scattering $fV\to f a$.

\begin{figure}
	\centering
	\includegraphics[width=0.49\columnwidth]{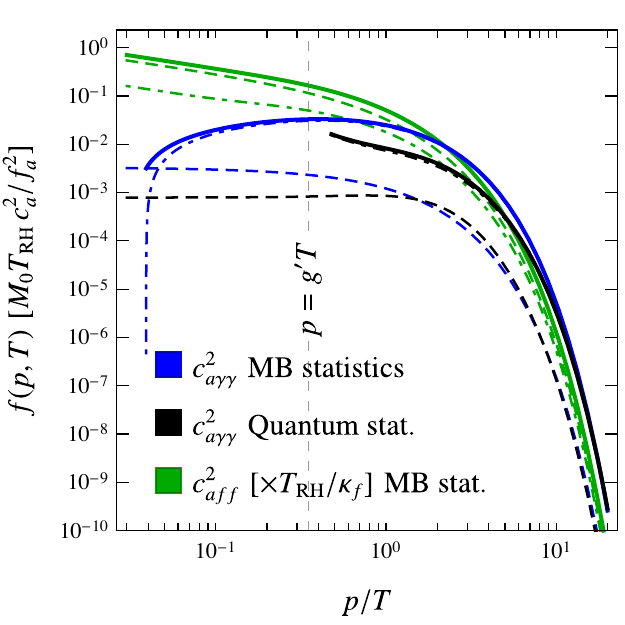} 
	\includegraphics[width=0.49\columnwidth]{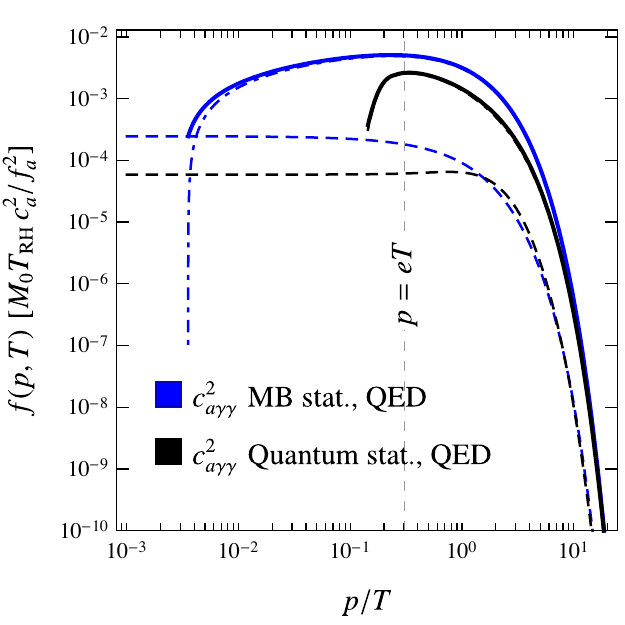} 
	\caption{ALP momentum distribution. \emph{Left panel:} In \emph{blue} we show the photophilic case for Maxwell-Boltzmann statistics; \emph{black} lines represent the photophilic case with quantum statistics and in \emph{green} we show the photophobic case for Maxwell-Boltzmann statistics.
For each, the dashed (dot-dashed) line represents $\bar{f}f\to V a$
($f V \to f a$) and the solid line is their sum. The vertical dashed line marks $p = g' T$, left of which thermal corrections should be added \cite{Braaten:1991dd}. \emph{Right panel:} Photophilic scenario for Maxwell-Boltzmann and quantum statistics in QED, namely, with only a single fermion with unit electric charge included.
	}
	\label{fig:ftot}
\end{figure}

The left panel of Figure~\ref{fig:ftot} shows the momentum distribution obtained for the photophilic (blue) and photophobic (green) ALP DM, in units of $M_0 T_\mathrm{RH} c_a^2/f_a^2$, where $c_a$ denotes $c_{a\gamma\gamma}$ or $c_{aff}$, respectively. The black lines correspond to the photophilic case with quantum statistics, and will be discussed later (see Sec.~\ref{subsec:quantum} below). The dashed lines represent the contribution from fermion annihilation, the dot-dashed ones correspond to fermion scattering, and solid lines are the sum.
We rescaled the photophobic ALP lines by $T_\mathrm{RH}/\kappa_f$, as otherwise, for $c_{a\gamma\gamma}\sim c_{aff}$ choice, green lines would not be visible on the same panel next to those corresponding to photophilic ALP. Moreover, since for the photophobic ALP $f(x)\propto m_f\, q^2 n_c$, with such rescaling the computed green curve is independent of the fermion choice.
It can be seen from the plot that, if both $c_{a\gamma\gamma}$ and $c_{aff}$ are present, the contribution from $c_{aff}$ is negligible as long as $c_{aff} \lesssim c_{a\gamma\gamma} T_\mathrm{RH}/m_f$.

Importantly, at $p/T \simeq 4\times 10^{-2}$ the dot-dashed blue line turns negative and hence the momentum distribution becomes unphysical for smaller values of $p/T$.
	  The reason for that is a simplified treatment of the IR divergence in the t-channel photon exchange diagram. For the purpose of our computations, we make a sharp cut at the location where the function turns negative. In the case of a Maxwell-Boltzmann distribution, we inferred this condition to be 
\begin{align}
 p/T = \text{Exp}\,[-1/4 + \gamma_E] \,\frac{m_{\gamma}^2}{4 T^2}\,.
\end{align} 
The same condition holds also for QED, with appropriately adjusted plasmon mass $m_\gamma$. 
Such single fermion scenario is shown in the right panel of \cref{fig:ftot}
and one can observe the shift of the cutoff toward an order of magnitude smaller values of $p/T$; this effect only arises due to different values for the plasmon mass in the full model and QED, respectively. In any case, for both panels, the cutoff occurs at a value of $p/T$ at least two orders of magnitude smaller than the expected mean, $\langle p \rangle /T$. If the cutoff was at $p/T\simeq \mathcal{O}(1)$, the calculation would not be justified. 

In passing, we mention that this effect has, to the best of our knowledge, not been discussed in the literature before. Typically, for UV freeze-in realizations where the production occurs through higher dimensional operators, the amplitude for the process is expressed only as a function of Mandelstam $s$ variable \cite{Elahi:2014fsa,Ballesteros:2020adh, DEramo:2020gpr} and in such cases there are clearly no problems with IR divergences. Here, we have, however, demonstrated that in a specific model, where t-channel diagram exists, full calculation should be performed and the impact of IR divergences should be assessed in order to have a control over consistency of the calculation.
We expect the unphysical low energy behaviour to disappear if a fully thermal treatment is adopted and postpone this to a future work.

Coming back to the left panel of \cref{fig:ftot}, one can infer that in the photophilic case, the annihilation process is subdominant, and as such it is often neglected in the literature. On the other hand, in the photophobic case it yields stronger contribution with respect to scattering (around a factor of 2, see also \cref{eq:omega-h2-1-photophobic,eq:omega-h2-2-photophobic}).
For the average momentum we obtained $\langle p \rangle/T = 3.24$ for a photophilic ALP and $\langle p \rangle/T = 2.36$ for a photophobic one. Up to higher order corrections, this result is independent of $m_f$.

\subsection{ALP distribution function computed using quantum statistics}\label{subsec:quantum}

The procedure described in Sec.~\ref{sec:MB stat} is only approximate, and it fails to capture two features that may \textit{a priori} be important. First, it cannot account correctly for the phase space regions where $p_i\lesssim T$, as in these regions the classical statistics deviates from the quantum one. Second, thermal field theory effects are important for $p_i \lesssim e T$, and must be correctly resummed. This is especially true for the process $f B\to f a$, which has a logarithmic IR divergence when $|\vec p_\gamma - \vec p_a|\to0$.
In that context, we are curious whether the usage of quantum statistics could in principle cure the occurrence of negative momentum distribution function at low $p/T$, as discussed in \cref{subsec:MB}. While we found there that the cutoff to be imposed is at values of $p/T$ that do not effectively impact $\langle p/T \rangle$, it would still be welcome to have physical values of momentum distribution function across all available momenta.
In order to properly take into account aforementioned features, we follow the procedure discussed in Ref.~\cite{Bolz:2000fu}.

In the photophilic scenario, and for the process $fB\to f a$, the collision term of Eq.~(\ref{eq:collision term}) (the ``hard" process) is evaluated by isolating the momentum flowing in the $t$-channel propagator $\vec k = \vec p_\gamma - \vec p_a$ and imposing an IR cutoff in the integration, $|\vec k|>k_\mathrm{cut}$. For $k_\mathrm{cut}\to0$ the integral diverges as $\log(T/k_\mathrm{cut})$.
This divergence is cured by adding a ``soft" term which is extracted from the ALP self-energy evaluated with a resummed photon propagator at finite temperature. It diverges as $\log(k_\mathrm{cut}/m_\gamma)$; therefore the sum is finite: $\log(T/m_\gamma)=-\log(\sqrt{11/12}g')$. The resulting collision term cannot be integrated fully analytically. The relevant expressions can be found in Eq.~(29) and Eq.~(21) of Ref.~\cite{Bolz:2000fu}, respectively for the hard and soft contribution (where the collision term, $\mathcal{C}$, is obtained by multiplying the interaction rate, $\Gamma$, by the Bose-Einstein distribution $1/(\exp(E_a/T)-1)$. We notice that these expressions are only accurate for $p_a > g' T$ (but they correctly account $p_{1,2,3}<g' T$). Below this value, the ALP-photon vertex should be thermally corrected as well~\cite{Braaten:1991dd}.

The collision term for the annihilation process $f \bar{f}\to\gamma a$ in the photophilic case can also be integrated with the full quantum statistics. By isolating the $s$-channel momentum $\vec q=\vec p_f + \vec p_{\bar{f}}$ and following a procedure similar to that of Ref.~\cite{Bolz:2000fu}, the integrals over $q$ and the angles can be computed analytically, while a numerical integration must be performed over the energies of the incoming particles. We did not include thermal field theory corrections in this case. We checked that, even with the full quantum statistics, the fermion annihilation process is subdominant compared to the scattering one, as we already found for the Maxwell-Boltzmann case (see again \cref{subsec:MB}).\\

The momentum distribution is obtained from the collision term by solving the Boltzmann equation, which is most conveniently written in terms of already introduced dimensionless quantities 
\begin{align}
\frac{m_H^2}{M_0 r} \frac{\partial}{\partial r} f(x,r) = \mathcal{C}(x,r) \,.
\label{eq:Boltzmann_equation}
\end{align}

The result is shown in black in the left panel of Fig.~\ref{fig:ftot}. The distribution is similar in shape compared to the one obtained with Maxwell-Boltzmann statistics. The overall normalization differs by a factor $\int d^3 p_a f_\mathrm{MB}/\int d^3 p_a f_\mathrm{QS}\sim5$; note that this is not relevant for the purpose of deriving structure formation limits.
The average momentum is, instead, very similar between two cases; we obtained $\langle p/T \rangle = 3.18$ for quantum statistics that is to be compared with $3.24$ from Maxwell-Boltzmann  (see \cref{subsec:MB}), yielding a $\approx 2\%$ difference. Due to practically identical results on $\langle p/T \rangle$ that stem from the two approaches, for the calculation of structure formation limits in \cref{sec:results}, we choose to proceed with the classical statistics description that allows us to work with fully analytical expressions, and leads to physical results even for lower momenta. We note that, as apparent from \cref{fig:ftot} (black lines), the change of the sign of the distribution function is not resolved with this procedure, and the dot-dashed lines come with a cutoff (see also right panel for QED). This is, however, not surprising because as stated above, the calculation is not valid for $p_a < g' T$. Interestingly, when quantum statistics is employed, the cutoff appears at somewhat larger values of momenta in comparison to Maxwell-Boltzmann case.

In passing, we briefly mention that the calculation of the collision term becomes more involved for a photophobic ALP. For each process, two Feynman diagrams have to be summed, in different channels (see Fig.~\ref{fig:feynman}). The interference between $s-$ and $t-$channel and between $t-$ and $u-$channel makes it impossible to define the exchanged momentum univocally, preventing an analytical integration of the collision term. For this scenario we content ourselves with the classical statistics results assuming that, as for the case of photophilic ALP, the quantum effects would affect very mildly the shape of the distribution and the average momentum $\langle p\rangle/T$.

 \section{Structure formation probes: Lyman-$\alpha$ and satellite counts}   
 \label{sec:data}
 
Generically, a DM candidate which features a non vanishing distribution function for rather large values of $p/T$ can be considered to be warm. As such, it starts to wash out structures at small scales and this can be quantified by the suppression of the
 matter power spectrum at such scales. \agree{The respective DM momentum distribution function $f(x,r)$ is essential in the computation of the matter power spectrum, $P(k)$ (the value of $k$ characterizes inverse scale): It explicitly enters in the calculation of the DM energy density fluctuation $\delta \rho$, which is needed to construct $P(k)$.} Hence, computing the matter power spectrum, is a starting point and the essence for assessing the structure formation limits%
 \footnote{A more precise estimation calls for $N$-body simulations approach which goes beyond the scope of this work.}. 

Qualitatively, this suppression can be quantified by the average momentum $\langle p/T \rangle$ of the respective ALP DM momentum distribution function corresponding to both production channels for which we find
\begin{align}
&\langle p/T \rangle_{\bar{f}f\to V a} = \frac{\int_0^\infty \dd x \, x^3 f_{\bar{f}f\to \gamma a}(x,r_\text{RH})}{\int_0^\infty \dd x \, x^2 f_{\bar{f}f\to \gamma a}(x,r_\text{RH})} = 3.0 \,\, (2.31)\,, \nonumber \\ 
&  \langle p/T \rangle_{f V \to f a} = \frac{\int_0^\infty \dd x \, x^3 f_{f \gamma \to f a}(x,r_\text{RH})}{\int_0^\infty \dd x \, x^2 f_{f \gamma \to f a}(x,r_\text{RH})} \approx 3.24 \,\, (2.44)\,, \nonumber \\ 
& \langle p/T \rangle = \frac{\int_0^\infty \dd x \, x^3 f(x,r_\text{RH})}{\int_0^\infty \dd x \, x^2 f(x,r_\text{RH})} \approx 3.24 \,\, (2.36)\,,
\label{eq:averaged_momentum_value}
\end{align}
where $r_\text{RH}=m_H/T_{\text{RH}}$; numbers in each line correspond to the photophilic and photophobic scnenario, the latter given in brackets. The last line refers to the sum of the two process, and it is the one which is physically relevant.

There are two complementary probes that we employ in our analysis: 
\emph{Ly-$\alpha$ forest} and the \emph{number of MW subhalos}. 
The former 
stands for a number of absorption lines occurring in the spectra of quasars and galaxies at higher redshift, stemming from the hydrogen in the inter-galactic medium \cite{Viel:2005qj, Narayanan:2000tp}. From such ``forests" one can obtain the matter power spectrum in one dimension, along the line of sight.

In the literature, it is customary to define a transfer function, $T(k)$,
\begin{align}
T(k) \equiv \sqrt{\frac{P(k)}{P(k)_{\Lambda\text{CDM}}}}\,,
\label{eq:transfer}
\end{align}
where subscript $\Lambda$CDM in the denominator denotes cold dark matter and the expression in the numerator is the power spectrum for ALP DM. In the following we calculate it \agree{numerically by feeding $f(x,r)$ into} \texttt{CLASS} \cite{Lesgourgues:2011re, Lesgourgues:2011rh}. \agree{Furthermore, we have to take into account entropy dilution effects due to a change in $g_*(T)$ between the time of production and today. This effect is quantified via an effective DM temperature $T_\mathrm{DM}$ which is related to the photon temperature $T_\gamma$ by
\begin{equation}
	T_\mathrm{DM} = \left( \frac{g_*(T_0)}{g_*(T_\mathrm{prod})} \right)^{1/3} T_\gamma = \left( \frac{3.94}{g_*(T_\mathrm{prod})} \right)^{1/3} T_\gamma.
	\label{eq:DM_temp}
\end{equation}
As can be seen this leads to a ``cooling" of the ALP DM and the strength of this effect depends on the time of production $T_\mathrm{prod}$. Above the electroweak phase transition we have $g_*(T) = 106.75$ for the SM.}

\begin{figure}
  \centering
  \includegraphics[width=0.75\columnwidth]{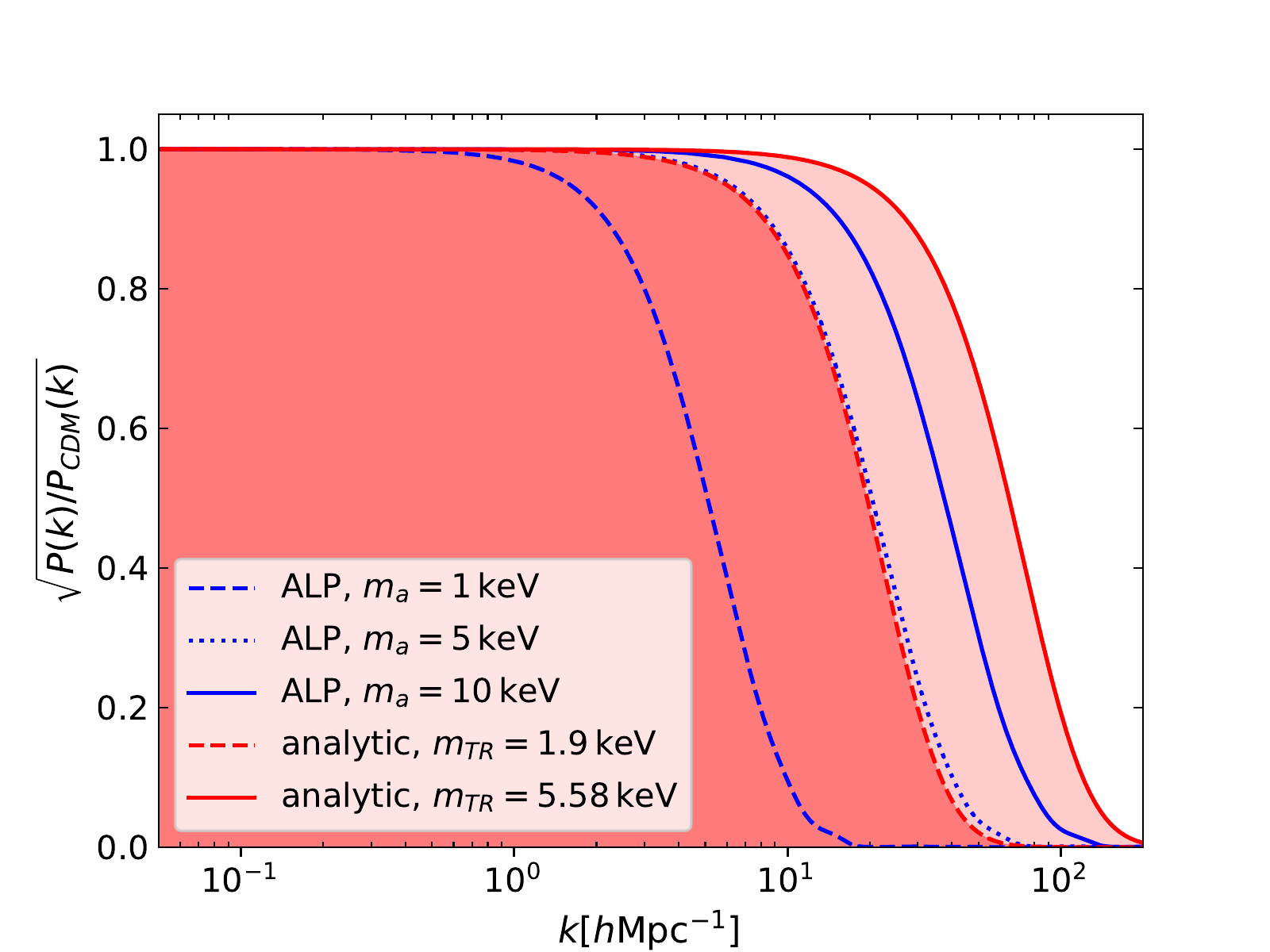}
  \caption{Transfer function (see \cref{eq:transfer}) shown for photophilic ALP DM with different masses compared to thermal relic limits presented as an analytic fit given in \cref{eq:fit}.}
  \label{fig:transfer}
\end{figure}

Then, the computed transfer function is compared to an analytical fit of the transfer function of a warm thermal relic with mass $m_\mathrm{TR}$ and abundance $\Omega_\mathrm{TR}$~\cite{Viel:2005qj}:
\begin{align}
T(k)= (1+(\alpha k)^{2\nu})^{-5/\nu}\,,
\label{eq:fit}
\end{align}
where $\nu=1.12$ and 
\begin{align}
\alpha = 0.049 \left( \frac{m_\text{TR}}{1 \, \text{keV}} \right)^{-1.11} \left( \frac{\Omega_\text{TR}}{0.25} \right)^{0.11} \left( \frac{h}{0.7} \right)^{1.22} \,h^{-1}\,\text{Mpc}\,.
\label{eq:transfer_func_ana}
\end{align}
Here, TR denotes a thermal relic which constitutes all of DM and for which structure formation limits are typically derived.
The values for the Hubble constant and $\Omega_\text{TR}$ are taken from \cite{Aghanim:2018eyx}.
The comparison is done by utilizing the half mode analyis \cite{Konig:2016dzg} where we define a scale, $k_{1/2}$, at which the squared transfer function has dropped to $0.5$, \textit{i.e.} $ T^2(k_{1/2}) = 0.5$. Then, for a given model and the corresponding transfer function, we check whether $T(k)^2 \geq T_\text{lim}(k)^2,\; \forall \, k \leq k_{1/2}$. If not, the considered parameter choice can be discarded. $T_\text{lim}$ is based on \cref{eq:fit} and uses
$m_\text{TR}$ adopted from dedicated searches using Ly-$\alpha$ forest observations.
The lower bounds on the mass of thermal relic, $m_{\text{TR}}$, span the range between $m_{\text{TR}}=1.9$ and $5.58\keV$ (see \cite{Garzilli:2019qki,Viel:2005qj,Yeche:2017upn,Irsic:2017ixq,Hsueh:2019ynk}). When showing our limits on ALP DM parameter space, we will present both conservative and aggressive bounds, corresponding to this mass range of $m_{\text{TR}}$.  
In \cref{fig:transfer} we show a comparison between \cref{eq:transfer} and \cref{eq:fit} for several values of ALP DM mass. One can infer that the more stringent limit ($m_\text{TR}=5.58\keV$) is essentially excluding 
ALP masses smaller than $\lesssim15\keV$, whereas by taking conservative limit of $m_\text{TR}=1.9\keV$, only $m_a \lesssim 5\keV$ are disfavored.

While we make use of the half mode analysis to infer limits on $m_a$ for a given $m_\text{TR}$, one can relate these two masses also by equalizing the free-streaming length, $\lambda_\text{FS}$, for both ALP and TR models. In the thermal relic case, $\lambda_{FS}\sim 0.22\,\text{Mpc} \left( \text{keV}/m_\text{TR} \right)^{4/3}$, whereas for the ALP, \agree{$\lambda_{FS}\sim \text{Mpc} \left( \text{keV}/m_a \right) \left( g_*(T_\nu)/g_*(T_\mathrm{prod}) \right)^{1/3}$ \cite{Heeck:2017xbu}}. Since we deal with \agree{large production temperatures, $T_\mathrm{prod}$}, we have \agree{$g{(T_\mathrm{prod})}=106.75$}, while $g_*(T_\nu)=10.75$. This allows one to derive the following relation between the two masses, $(m_a)_\mathrm{min} \simeq 2.1\keV \left( \frac{m_\text{TR}}{1\keV} \right)^{4/3}$; a similar result was quoted in \cite{Bae:2017dpt}. A more appropriate derivation which takes the different momentum distributions into account is to use the half mode scale $k_{1/2}$. Equalizing it for both models we found the following numerical relation
\begin{align}
	(m_a)_\mathrm{min} \simeq 2.1\keV \left( \frac{m_\text{TR}}{1\keV} \right)^{1.27}.
\end{align}
\agree{Assuming a different value for $g_*(T\mathrm{prod})$ one has to modify this relation by taking the entropy dilution (see \cref{eq:DM_temp}) into account. As such, the previous equation is modified by an additional factor
\begin{align}
(m_a)_\mathrm{min} \simeq 2.1\keV \left( \frac{m_\text{TR}}{1\keV} \right)^{1.27} \left( \frac{106.75}{g_*(T_\mathrm{prod})} \right)^{1/3}.
\label{eq:alp_tr_mass_relation}
\end{align}
Therefore, smaller values of $g_*(T_\mathrm{prod})$ have to be compensated by larger ALP DM masses. We will come back to this when discussing limits for the photophobic scenario.	}



\medskip

As far as the number of MW satellites is concerned, ALP DM with 
momentum distribution as in \cref{eq:averaged_momentum_value}
would cause the occurrence of smaller number of subhalos, compared to $\Lambda$CDM. The predicted number of satellites has an intriguing connection with the type of DM and its production;  we are able to set the limit on $m_a$ by requiring that the number of satellites is not smaller that what was observed. For counting  subhalos we follow the approach outlined in \cite{Schneider:2016uqi,Murgia:2017lwo}: 11 ``classical" satellites are combined with 15 ultra-faint satellites found by SDSS. Latter number is multiplied by a factor of $7/2$ because of the limited sky coverage of SDSS; this yields in total $N_\text{sub}=64$. One should note that, in addition to SDSS, several more ultra-faints satellites or candidates have at least been reported (see for instance \cite{Drlica-Wagner:2019vah}) by other surveys, so we regard $N_\text{sub}=64$ as a conservative estimate for the number of MW companions.

We follow the approach from \cite{Schneider:2013ria,Schneider:2014rda,Schneider:2016uqi} where the authors present a formula to derive $N_\text{sub}$ for a given matter power spectrum, based on an extended Press-Schechter approach
\begin{align}
\frac{\dd N_{sub}}{\dd M_{sub}}= \frac{1}{C} \frac{1}{6\pi^2} \frac{M_{MW}}{M_{sub}^2} \frac{P(1/R_{sub})}{R_{sub}^3\sqrt{2\pi(S_{sub}-S_{MW})}}\,.
\label{eq:number_sub_halos}
\end{align}
Here, $R_{sub}$ is the radius of a subhalo that has a mass $M_{sub}$; these two quantities are related through $M_{sub} = \frac{4\pi}{3} \Omega_m \rho_c (2.5R_{sub})^3$. The power spectrum $P(k)$ of the warm ALP DM has to be integrated for scales smaller compared to the respective subhalo or MW scale 
\begin{align}
	S_i=\frac{1}{2\pi^2} \int_0^{R_i^{-1}}\dd k \, k^2 P(k)\,.
	\label{eq:sven}
\end{align}
Decreasing $m_a$ leads to smaller variances $S_i$ (smaller denominator in \cref{eq:number_sub_halos}) as well as a suppression of $P(k)$ which then features an earlier drop (smaller numerator in \cref{eq:number_sub_halos}); the latter effect is, however, stronger and therefore less numbers of subhalos $N_{sub}$ are predicted.
 
We further note here that the limits clearly depend on the MW mass, $M_{MW}$.
The precise value of this quantity is still under investigation and, depending on the analysis, the reported values range roughly between $1\times 10^{12}\, M_\odot <M_{MW}<1.5\times 10^{12}\,M_\odot$ (see \cite{Wang:2015ala,Callingham:2018vcf,Cautun:2019eaf,Karukes:2019jwa,Wang:2019ubx} and references therein), using recent data from the GAIA survey. In the following, we will refer to the lower mass as \textit{aggressive}, while the higher mass bound is dubbed as \textit{conservative}. We want to stress the following two subtleties: First, the quoted MW masses are defined with respect to densities $200$ times larger than the critical density, the so called virial mass of the MW. As such, the constant $C$ of \cref{eq:number_sub_halos} is given by $C=34$. Second, \cref{eq:number_sub_halos} uses MW masses in units of $M_\odot /h$ as an input and one has to take this additional factor into account.


\section{Suppressing ALP production from misalignment and topological defects}
\label{sec:discussion}

An important assumption of our discussion is that ALP is predominantly produced through a freeze-in process. This means that the misalignment contribution, as well as any ALP population produced from the decay of topological defects such as cosmic strings, must be suppressed. \agree{As we are going to show in what follows,  this is not trivial for the photophilic ALP,  as the requirement of a small misalignment energy density implies bounds on $T_\mathrm{RH}$ and $f_a$ which are in contrast with what is needed in order to match the observed DM abundance with the freeze-in population. 
For this reason,  in order to justify the assumption that the misalignment energy density is suppressed, we will need some specific assumption about the axion potential and the specific symmetry breaking pattern through which it develops in the early universe. 
This statement holds only for the photophilic axion,  as we will detail below.}

The energy density from ALP misalignment, assuming a constant ALP mass, is \cite{Arias:2012az}
\begin{align}\label{eq:rhoa0}
\rho_{a,0} = 0.17 \,\frac{\mathrm{keV}}{\mathrm{cm}^3} \left(\frac{m_a}{\mathrm{eV}}\right)^{1/2}\left(\frac{a_i}{10^{11}\,\mathrm{GeV}}\right)^2 \,.
\end{align}
Requiring that this value does not exceed the observed DM abundance yields
\begin{equation}
\label{eq:max angle}
a_i < 2.6 \times 10^{10}\,\mathrm{GeV} \left(\frac{10\,\mathrm{keV}}{m_a}\right)^{1/4} \,,
\end{equation}
where $a_i$ is the displacement from the minimum when ALP starts to oscillate. If the ALP is present during inflation\footnote{If the ALP is the pseudo-Goldstone boson of a spontaneously broken global symmetry, this corresponds to the scenario where the symmetry is broken before or during inflation\agree{, with $H_I<f_a$}.
In any case, this assumption is not crucial for our discussion.},
 the minimal misalignment is set by quantum fluctuations during inflation, $a_i > \sqrt{N_e} \frac{H_I}{2\pi} $. Here, $N_e$ is the number of efolds of inflation and $H_I$ is the Hubble rate during inflation. This, combined with Eq.~(\ref{eq:max angle}), results in a rather stringent upper limit for $H_I$. Recalling that the maximal reheating temperature is obtained under the assumption of instantaneous reheating (\textit{i.e.} the case in which the entire energy density of inflation is converted into radiation), we obtain a bound on the reheating temperature
\begin{align}
T_\mathrm{RH} < 1.2 \times 10^{14} \,\mathrm{GeV} \left(\frac{106.75}{g_*}\right)^{1/4} \left(\frac{10\,\mathrm{keV}}{m_a}\right)^{1/8} \left(\frac{60}{N_e}\right)^{1/4}  \,.
\label{eq:TRH}
\end{align}
With such values of $T_\text{RH}$ in the photophilic scenario, matching the DM abundance would typically be achievable for large $c_{a\gamma\gamma}/f_a$ couplings, disfavored by X-ray constraints (see $T_{\text{RH}}=10^{14} \, \text{GeV}$ line in Fig.~\ref{fig:photophilic}).
\agree{For the photophobic scenario,  instead,  this bound is always satisfied in our parameter space.}

\agree{Let us now present two alternative ways in which the misalignment contribution can be suppressed without implying a low reheating temperature as in Eq.~(\ref{eq:TRH}).}\\

 If the Peccei-Quinn symmetry breaks after inflation,
 the misalignment angle, neglecting anharmonicities in the potential, is averaged to $\langle a_i^2 \rangle/f_a^2 = \pi^2/3$. After inserting this value into Eq.~(\ref{eq:rhoa0}) and imposing $\rho_{a,0} < \rho_{\mathrm{CDM},0}$ we obtain
\begin{align}\label{eq:limit f misalignment after inflation}
f_a < 1.4\times 10^{10}\,\mathrm{GeV} \left(\frac{10\,\mathrm{keV}}{m_a}\right)^{1/4} \,.
\end{align}

In this case
we have no upper bound on the reheating temperature. We only need to assume that the ALP-SM couplings are suppressed compared to $f_a$, \textit{i.e.} $c_{a\gamma\gamma}, c_{aff}\ll \alpha/(2\pi)$, the RHS being the reference value of this coupling in typical axion models.
Further, one needs to make sure that the contribution from the decay of cosmic string is negligible.
For the typical example of a pNGB from the breaking of a global $U(1)$ symmetry, such as the standard QCD axion, the axion density from the decay of topological defects is similar in magnitude to the one from misalignment. Assuming this is the case, we expect both contributions to be subdominant when Eq.~(\ref{eq:limit f misalignment after inflation}) is satisfied.\\

The second way of suppressing the misalignment energy density is to have a very small angle from the start. This is achieved if Peccei-Quinn symmetry is broken already during inflation and the axion is heavy, $m_a > H_I$. If the axion mass is constant, this would imply a too strong upper bound on $T_\mathrm{RH}$. Still, the mass of the axion could have been much larger during inflation than today (see \textit{e.g.}~\cite{Dvali:1995ce, Co:2018phi} and references therein for some explicit realizations). In particular, for $m_a\approx 10^{16}\,\mathrm{GeV}$, the reheating temperature can be as large as $10^{17}\,\mathrm{GeV}$ during inflation.
If, after inflation, the ALP is light again, thermal fluctuations could in principle increase the energy density from the misalignment mechanism. This is, however, not the case for the photophilic scenario under our consideration as ALPs do not thermalize with the SM plasma.

\agree{
We again stress that, for the photophobic ALP, the requirement of a vanishing misalignment energy density is easy to satisfy. If the axion is present during inflation, Eq.~(\ref{eq:TRH}) applies, which is always satisfied for our choice $T_\mathrm{RH}\lesssim\mathcal{O}(100 \mathrm{GeV})$. If, instead, the PQ symmetry is broken after inflation,  $f_a$ should satisfy Eq.~(\ref{eq:limit f misalignment after inflation}). Comparing this requirement with the preferred value of $c_{aff}/f_a$ shown in Fig.~\ref{fig:photophobic}, obtained by imposing that the freeze-in abundance matches the observed DM one, results in $c_{aff}\lesssim \mathcal{O}(1)$.
}


\section{Results}
\label{sec:results}
We use the distribution functions in \cref{eq:f photophilic Teresi,eq:f photophobic Teresi} to calculate the matter power spectrum using \texttt{CLASS}, derive the corresponding transfer function and the number of MW subhalos and finally compare against results from observations following the strategy outlined in \cref{sec:data}.
As long as the observed DM abundance is matched, the result does not depend on the couplings $c_{a\gamma \gamma}$ and $c_{aff}$; hence, we performed a scan over $m_a$ to derive limits.

\begin{figure}
	\centering
	\includegraphics[width=0.7\columnwidth]{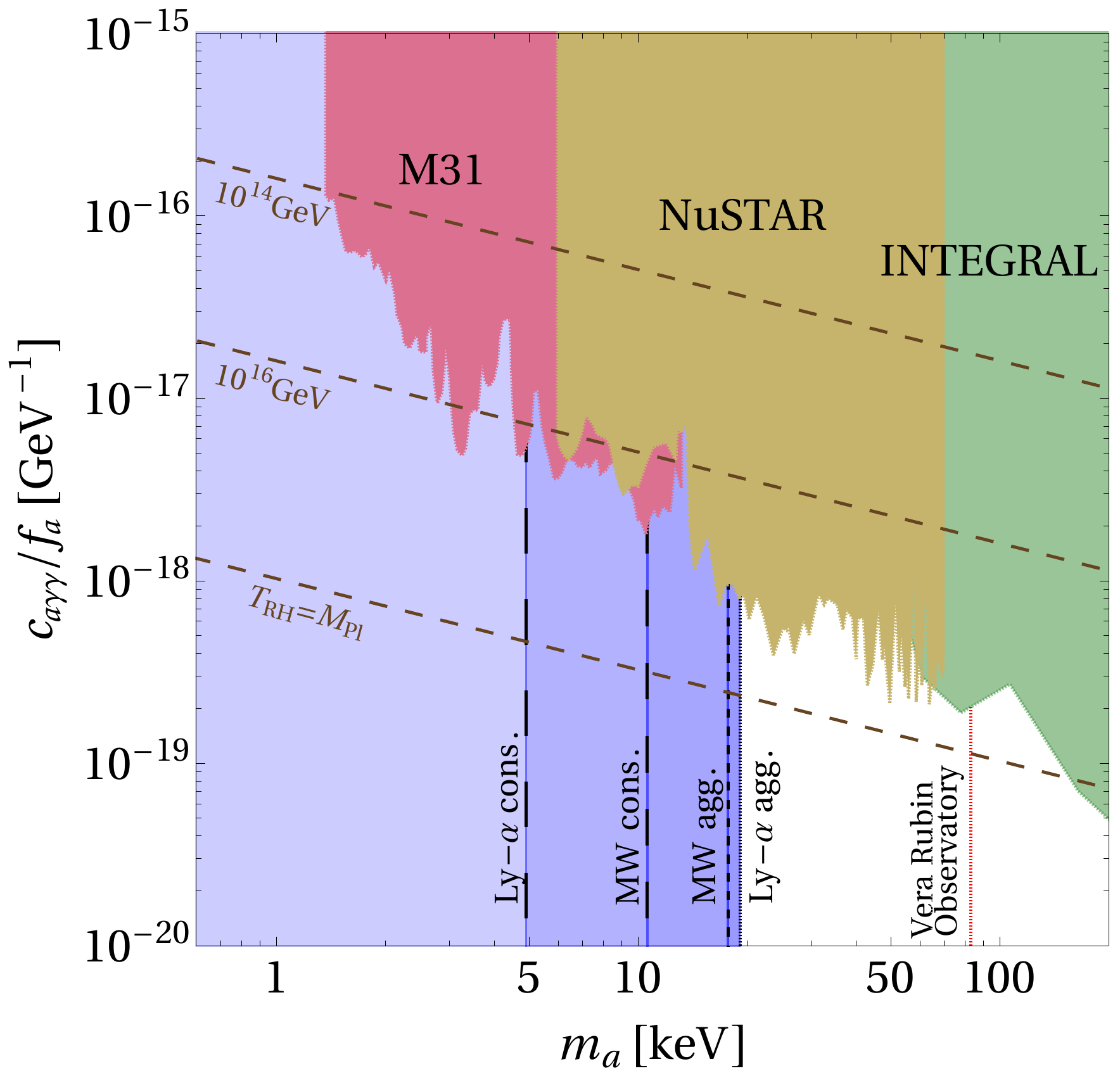}
	\caption{Structure formation limits on the \emph{photophilic} ALP parameter space derived using the half mode analysis technique as well as MW subhalo count. The former (latter) are denoted as $Ly-\alpha$ (MW). For both, conservative and aggressive limits are shown (see \cref{sec:data} for details). What is also shown are X-ray limits from INTEGRAL \cite{Boyarsky:2007ge}, NuSTAR \cite{Roach:2019ctw} and M31 \cite{Horiuchi:2013noa}. The diagonal dashed lines indicate two specific values of $T_\text{RH}$, calculated by the requirement of producing the observed amount of DM. Sensitivity projection at $m_a \sim 80$ keV from forthcoming Vera Rubin observatory is shown with red line.}
	\label{fig:photophilic}
\end{figure}

The results for the Ly-$\alpha$ forest and the MW subhalo counts for the photophilic scenario are shown in \cref{fig:photophilic}. For both probes, we show conservative and aggressive bounds based on the arguments presented in \cref{sec:data}. The explicit values for the lower limits on the DM mass $m_a$ are also quoted in \cref{tab:1}. \agree{For this scenario we set $g_*(T_\mathrm{prod}) = 106.75$ for all parameter choices because of the high reheating scale involved.}

\begin{table}[h!]
	\centering
	\begin{tabular}{c|c|c}
						&	\emph{cons.}	&	\emph{agg.} \\
			\hline
		Lyman-$\alpha$	&	$4.9\keV$	&	$19.1\keV$ \\
		MW subhalo		&	$10.3\keV$	&	$17.4\keV$
	\end{tabular}
	\caption{Structure formation limits in the \emph{photophilic} scenario.}
	\label{tab:1}
\end{table}
 
 For this photophilic scenario we also superimpose limits from X-ray searches which severely narrow the viable parameter space in the $m_a \gtrsim 17\keV$ region where structure formation constraints fade away. We have calculated X-ray limits on ALP DM by utilizing existing ones on keV-scale sterile neutrino DM  \cite{Adhikari:2016bei} as well as the expression for photon fluxes in both models. 
 

\textcolor{black}{The diagonal lines indicate the parameter space for which reheating temperature equals $T_{\text{RH}} = 10^{14}\GeV$,  $10^{16}\GeV$ and the reduced Planck mass $M_{\text{Pl}}$,  respectively. The middle one corresponds to the upper bound on the energy scale during inflation set by Planck through the tensor-to-scalar ratio $r$~\cite{Planck:2018jri}. If there are scenarios in which the limit can be avoided, the maximal allowed reheating temperature would correspond to the reduced
Planck mass (lowest line in the figure). As indicated by the plot, most of the parameter space for a photophilic ALP which is not constrained by structure formation is already probed by X-ray searches.
More available parameter space can be achieved by considering additional ALP couplings to $W^a$ bosons, which allow for reheating temperatures smaller by a factor $\approx 1/20$ due to the larger gauge coupling and to the different trace factors. Namely, ALP coupling $\sim 5$ times smaller with respect to the axion-photon interaction scenario, suffices for producing the observed amount of dark matter. Finally, we want to stress that structure formation limits on the photophilic ALP are exceeding X-ray bounds for $m_a \lesssim$ 2 keV independently of the reheating temperature.}


We should briefly discuss the previously reported unidentified line at $\sim3.5\keV$ in a X-ray data spectrum \cite{Bulbul:2014sua,Boyarsky:2014jta}. While such discovery has chiefly received explanations in terms of decaying keV-scale sterile neutrino DM, $7\keV$ ALP DM 
was also discussed \cite{Jaeckel:2014qea}. Our aggressive structure formation limits are, however, clearly disfavoring such an interpretation, while conservative ones are marginally consistent with it. \agree{We should still stress that our findings are not in general disfavoring a DM interpretation of a $3.5\keV$ line and hold only for the freeze-in production of ALPs. For instance, scenarios where ALPs are dominantly produced via the misalignment mechanism are still viable in this regard. 
}

\begin{figure}
	\centering
	\includegraphics[width=0.7\columnwidth]{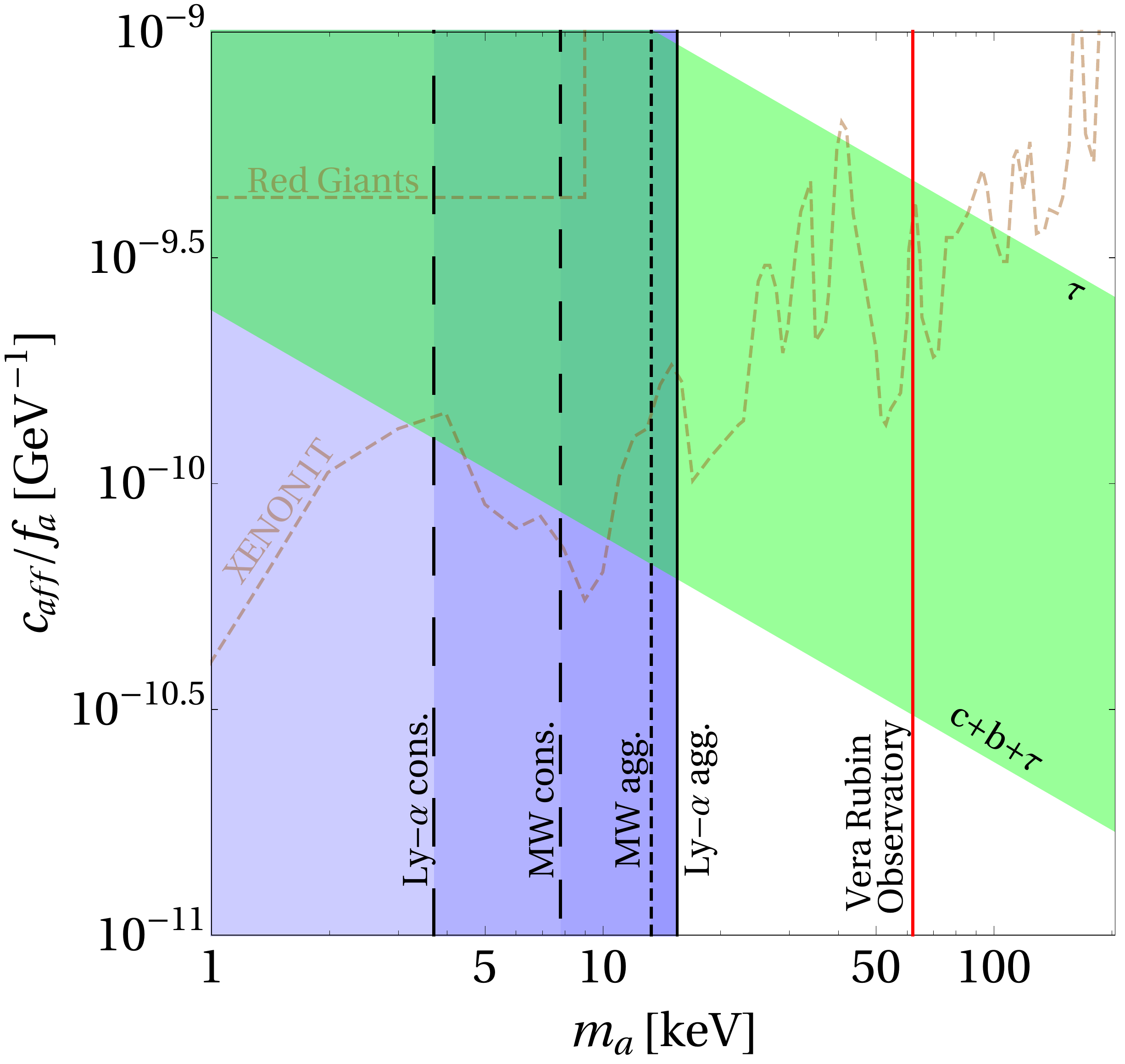}
	\caption{Structure formation limits on the \emph{photophobic} ALP parameter space derived using the half mode analysis technique as well as MW subhalo count. The former (latter) are denoted as $Ly-\alpha$ (MW). For both, conservative and aggressive limits are shown (see \cref{sec:data} for details). Theoretically viable parameter space is shown as a green diagonal band, spanning cases in which ALP couples only to $\tau$-lepton (upper edge) and where it interacts with $\tau$-lepton, c- and b-quark (lower edge). Sensitivity projection at $m_a \sim 60$ keV from forthcoming Vera Rubin observatory is shown with red line.}
	\label{fig:photophobic}
\end{figure}

\medskip

Next, we apply the same techniques to the photophobic scenario; in that case we derived \cref{eq:f photophobic Teresi} by expanding \cref{eq:master_eq_momentum_distr} for small $m_f$ assuming that $T_\text{RH} \gg m_f$. By doing so, the dependence on the reheating temperature actually drops out\footnote{We implicitly assumed that $T_\text{RH}\lesssim 160\GeV$ and this approximation holds up to $T_\text{RH} \geq 10\GeV$ or so. For smaller reheating temperatures one has to keep the full expression and an explicit $T_\text{RH}$ dependence is recovered.}.
The accessible parameter space is then represented by a linear relation between $m_a$ and $c_{aff}$. The results are shown in \cref{fig:photophobic} where the green diagonal band  denotes the viable parameter space of the model. The upper edge of the band corresponds to a $\tau$-lepton contribution only, while the lower edge shows the sum of $\tau$-lepton, c- and b-quark contributions, with equal coefficients.

 The regions shown in blue are disfavored using Ly-$\alpha$ forest and the MW subhalo counts. The most aggressive limit (see also \cref{tab:2}) reads $m_a\approx 15\keV$ and we hence predict that only the parameter space above such values could be realizable in Nature. The derived constraints are slightly weaker compared to the photophilic case, because ALP DM produced in the photophobic scenario is ``cooler" compared to the other case\footnote{A recent study on updated numbers of MW satellites quotes results in terms of $m_\text{TR}>6.5\keV$~\cite{Nadler:2020prv}. As such we can use the ALP DM transfer function directly to derive mass bounds; after doing so we found $m_a>23.0\,\keV$ for photophilic and $m_a>17.5\,\keV$ for the photophobic scenario, which extends even beyond \emph{aggressive} bounds derived in our own analysis.} (see \cref{sec:model}). \agree{However, in the photophobic scenario we assumed $T_\mathrm{RH}\lesssim 160\GeV$ and so one has to take a change in $g_*(T_\mathrm{prod})$ into account. The results were derived with $g_*(T_\mathrm{prod}) = 106.75$ and so the limits on the ALP mass have to be rescaled according to \cref{eq:alp_tr_mass_relation}. For instance, if $g_*(T_\mathrm{prod}) = 80$, all mass limits are increased by a factor $\simeq 1.1$ compared to the results shown in \cref{tab:2}. }
 
 \begin{table}[h!]
	\centering
	\begin{tabular}{c|c|c}
		&	\emph{cons.}	&	\emph{agg.} \\
		\hline
		Lyman-$\alpha$	&	$3.7\keV$	&	$15.5\keV$ \\
		MW subhalo		&	$7.8\keV$	&	$13.3\keV$
	\end{tabular}
	\caption{Structure formation limits in the \emph{photophobic} scenario.}
	\label{tab:2}
\end{table}

If the ALP has flavor universal couplings or if the strength of ALP  couplings to fermions is at least comparable across several flavors including electrons, then the red giants \cite{Viaux:2013lha, Bertolami:2014wua} and XENON1T \cite{Aprile:2020tmw} limits, shown on \cref{fig:photophobic}, are applicable since they constrain the ALP-electron coupling. The latter is clearly more stringent and relevant for larger span of ALP masses, 
excluding $c_{aee}/f_a\gtrsim 10^{-10}\GeV^{-1}$. Clearly, the XENON1T line is chiefly below the green band in the region $m_a \lesssim 10$ keV, disfavoring the flavor universal ALP coupling scenario for such parameter choice. 

We have also evaluated constraints from X-ray searches using expressions from \cite{Craig:2018kne}. Namely, if ALP couples to fermions, interaction with photons will be induced at the quantum level. We have, however, found that such X-ray limits are only relevant if ALP couples to electrons, because the loop induced coupling to photons is suppressed by the mass ratio $m_a^2/m_f^2$. In such case, even parameter space above $m_a\sim 10$ keV, unexcluded by XENON1T, would be disfavored in flavor universal scenario. We note, however, that it was recently shown that by adding more new physics, destructive interference between loop-level diagrams contributing to decays into SM photons can be achieved; hence the limit can be relaxed substantially \cite{Benso:2019jog}, particularly for such higher values of $m_a$.

We note here that our results constrain the relaxion DM model of Ref.~\cite{Fonseca:2018kqf}, \agree{in which relaxion production occurs via freeze-in}. The relaxion can constitute 100\% of DM only if a large mass is allowed by a large hierarchy in the relaxion couplings.
Even assuming the most conservative bounds from the number of MW subhalos, the model provides an explanation of the XENON1T anomaly only if the relaxion is responsible for a fraction of the DM abundance~\cite{Fonseca:2020pjs}.\\

Finally, let us address future sensitivity projections. Utilizing data from the upcoming Vera C. Rubin observatory, designed to measure MW mass halo function down to $10^6 \,M_\odot$, it will be 
possible to improve limits dramatically.
Using its projected thermal mass limit  $m_\text{TR}>18\keV$ \cite{Drlica-Wagner:2019xan,Dvorkin:2020xga} we find that, for the photophilic case, the lower limit on the ALP DM mass would be pushed to $m_a>83\keV$. Similarly, we find $m_a > 62\keV$ for the photophobic case (both values are illustrated in \cref{fig:photophilic,fig:photophobic} with red lines). It is rather intriguing that for the case of photophilic ALP, the structure formation limits are expected to start competing with the long-standing X-ray limits in $m_a \sim \mathcal{O}(100)$ keV region.



\section{Summary and Conclusion} 
\label{sec:conclude}

ALPs are currently one of the most popular beyond the SM extensions, being studied and tested across several mass scales. 
In this paper we studied keV-scale ALP DM produced via freeze-in
through feeble interactions with photons and SM fermions.
The respective momentum distribution has been calculated assuming Maxwell-Boltzmann and quantum statistics. 
Although these approaches do not capture the full picture at small values of $p/T$, they turn out to be robust for constraining the parameter space using the matter power spectrum derived from the respective momentum distribution function. 
Using Lyman-$\alpha$ forest data as well as the observed number of MW companions we derived structure formation limits that were missing to date. For the photophilic ALP DM, we found the most aggressive limits to exclude ALP DM masses below 
$\sim 19\keV$, complementing constraints from X-ray data.  
For photophobic ALP, the obtained limits are somewhat milder, because in that case we found DM to be ``cooler". Utilizing measurements from the upcoming Vera Rubin observatory, such bounds will be strongly improved to $m_a \gtrsim 60\keV$ for both scenarios.


\section*{Acknowledgements}
We would like to thank Joerg Jaeckel for discussions. 
The work of SB and EM is supported by the Cluster of Excellence ``Precision Physics, Fundamental Interactions, and Structure of Matter'' (PRISMA+ EXC 2118/1) funded by  the  German  Research  Foundation (DFG)  within  the German  Excellence  Strategy  (Project  ID  39083149).
Fermilab is operated by the Fermi Research Alliance, LLC under contract No. DE-AC02-07CH11359 with the United States Department of Energy.


\bibliographystyle{JHEP}
\bibliography{refs}

\providecommand{\href}[2]{#2}\begingroup\raggedright\begin{thebibliography}{10}

\bibitem{Arcadi:2017kky}
G.~Arcadi, M.~Dutra, P.~Ghosh, M.~Lindner, Y.~Mambrini, M.~Pierre, S.~Profumo,
  and F.~S. Queiroz, {\it {The waning of the WIMP? A review of models,
  searches, and constraints}},  {\em Eur. Phys. J. C} {\bf 78} (2018), no.~3
  203, [\href{http://arxiv.org/abs/1703.07364}{{\tt arXiv:1703.07364}}].

\bibitem{Dodelson:1993je}
S.~Dodelson and L.~M. Widrow, {\it {Sterile-neutrinos as dark matter}},  {\em
  Phys. Rev. Lett.} {\bf 72} (1994) 17--20,
  [\href{http://arxiv.org/abs/hep-ph/9303287}{{\tt hep-ph/9303287}}].

\bibitem{Shi:1998km}
X.-D. Shi and G.~M. Fuller, {\it {A New dark matter candidate: Nonthermal
  sterile neutrinos}},  {\em Phys. Rev. Lett.} {\bf 82} (1999) 2832--2835,
  [\href{http://arxiv.org/abs/astro-ph/9810076}{{\tt astro-ph/9810076}}].

\bibitem{Merle:2013wta}
A.~Merle, V.~Niro, and D.~Schmidt, {\it {New Production Mechanism for keV
  Sterile Neutrino Dark Matter by Decays of Frozen-In Scalars}},  {\em JCAP}
  {\bf 03} (2014) 028, [\href{http://arxiv.org/abs/1306.3996}{{\tt
  arXiv:1306.3996}}].

\bibitem{Brdar:2017wgy}
V.~Brdar, J.~Kopp, J.~Liu, and X.-P. Wang, {\it {X-Ray Lines from Dark Matter
  Annihilation at the keV Scale}},  {\em Phys. Rev. Lett.} {\bf 120} (2018),
  no.~6 061301, [\href{http://arxiv.org/abs/1710.02146}{{\tt
  arXiv:1710.02146}}].

\bibitem{deGouvea:2019phk}
A.~De~Gouv\^ea, M.~Sen, W.~Tangarife, and Y.~Zhang, {\it {Dodelson-Widrow
  Mechanism in the Presence of Self-Interacting Neutrinos}},  {\em Phys. Rev.
  Lett.} {\bf 124} (2020), no.~8 081802,
  [\href{http://arxiv.org/abs/1910.04901}{{\tt arXiv:1910.04901}}].

\bibitem{Lee:2017qve}
J.-W. Lee, {\it {Brief History of Ultra-light Scalar Dark Matter Models}},
  {\em EPJ Web Conf.} {\bf 168} (2018) 06005,
  [\href{http://arxiv.org/abs/1704.05057}{{\tt arXiv:1704.05057}}].

\bibitem{Hui:2016ltb}
L.~Hui, J.~P. Ostriker, S.~Tremaine, and E.~Witten, {\it {Ultralight scalars as
  cosmological dark matter}},  {\em Phys. Rev. D} {\bf 95} (2017), no.~4
  043541, [\href{http://arxiv.org/abs/1610.08297}{{\tt arXiv:1610.08297}}].

\bibitem{Redondo:2008ec}
J.~Redondo and M.~Postma, {\it {Massive hidden photons as lukewarm dark
  matter}},  {\em JCAP} {\bf 02} (2009) 005,
  [\href{http://arxiv.org/abs/0811.0326}{{\tt arXiv:0811.0326}}].

\bibitem{Hambye:2019dwd}
T.~Hambye, M.~H. Tytgat, J.~Vandecasteele, and L.~Vanderheyden, {\it {Dark
  matter from dark photons: a taxonomy of dark matter production}},  {\em Phys.
  Rev. D} {\bf 100} (2019), no.~9 095018,
  [\href{http://arxiv.org/abs/1908.09864}{{\tt arXiv:1908.09864}}].

\bibitem{Arias:2012az}
P.~Arias, D.~Cadamuro, M.~Goodsell, J.~Jaeckel, J.~Redondo, and A.~Ringwald,
  {\it {WISPy Cold Dark Matter}},  {\em JCAP} {\bf 06} (2012) 013,
  [\href{http://arxiv.org/abs/1201.5902}{{\tt arXiv:1201.5902}}].

\bibitem{Jaeckel:2014qea}
J.~Jaeckel, J.~Redondo, and A.~Ringwald, {\it {3.55 keV hint for decaying
  axionlike particle dark matter}},  {\em Phys. Rev. D} {\bf 89} (2014) 103511,
  [\href{http://arxiv.org/abs/1402.7335}{{\tt arXiv:1402.7335}}].

\bibitem{Im:2019iwd}
S.~H. Im and K.~S. Jeong, {\it {Freeze-in Axion-like Dark Matter}},  {\em Phys.
  Lett. B} {\bf 799} (2019) 135044,
  [\href{http://arxiv.org/abs/1907.07383}{{\tt arXiv:1907.07383}}].

\bibitem{Ouellet:2018beu}
J.~L. Ouellet et~al., {\it {First Results from ABRACADABRA-10 cm: A Search for
  Sub-$\mu$eV Axion Dark Matter}},  {\em Phys. Rev. Lett.} {\bf 122} (2019),
  no.~12 121802, [\href{http://arxiv.org/abs/1810.12257}{{\tt
  arXiv:1810.12257}}].

\bibitem{Marsh:2013ywa}
D.~J.~E. Marsh and J.~Silk, {\it {A Model For Halo Formation With Axion Mixed
  Dark Matter}},  {\em Mon. Not. Roy. Astron. Soc.} {\bf 437} (2014), no.~3
  2652--2663, [\href{http://arxiv.org/abs/1307.1705}{{\tt arXiv:1307.1705}}].

\bibitem{Fonseca:2018kqf}
N.~Fonseca and E.~Morgante, {\it {Relaxion Dark Matter}},  {\em Phys. Rev. D}
  {\bf 100} (2019), no.~5 055010, [\href{http://arxiv.org/abs/1809.04534}{{\tt
  arXiv:1809.04534}}].

\bibitem{Fonseca:2020pjs}
N.~Fonseca and E.~Morgante, {\it {Probing photophobic (rel)axion dark matter}},
   \href{http://arxiv.org/abs/2009.10974}{{\tt arXiv:2009.10974}}.

\bibitem{Banerjee:2018xmn}
A.~Banerjee, H.~Kim, and G.~Perez, {\it {Coherent relaxion dark matter}},  {\em
  Phys. Rev. D} {\bf 100} (2019), no.~11 115026,
  [\href{http://arxiv.org/abs/1810.01889}{{\tt arXiv:1810.01889}}].

\bibitem{Bulbul:2014sua}
E.~Bulbul, M.~Markevitch, A.~Foster, R.~K. Smith, M.~Loewenstein, and S.~W.
  Randall, {\it {Detection of An Unidentified Emission Line in the Stacked
  X-ray spectrum of Galaxy Clusters}},  {\em Astrophys. J.} {\bf 789} (2014)
  13, [\href{http://arxiv.org/abs/1402.2301}{{\tt arXiv:1402.2301}}].

\bibitem{Boyarsky:2014jta}
A.~Boyarsky, O.~Ruchayskiy, D.~Iakubovskyi, and J.~Franse, {\it {Unidentified
  Line in X-Ray Spectra of the Andromeda Galaxy and Perseus Galaxy Cluster}},
  {\em Phys. Rev. Lett.} {\bf 113} (2014) 251301,
  [\href{http://arxiv.org/abs/1402.4119}{{\tt arXiv:1402.4119}}].

\bibitem{Aprile:2020tmw}
{\bf XENON} Collaboration, E.~Aprile et~al., {\it {Excess electronic recoil
  events in XENON1T}},  {\em Phys. Rev. D} {\bf 102} (2020), no.~7 072004,
  [\href{http://arxiv.org/abs/2006.09721}{{\tt arXiv:2006.09721}}].

\bibitem{Merle:2015vzu}
A.~Merle, A.~Schneider, and M.~Totzauer, {\it {Dodelson-Widrow Production of
  Sterile Neutrino Dark Matter with Non-Trivial Initial Abundance}},  {\em
  JCAP} {\bf 04} (2016) 003, [\href{http://arxiv.org/abs/1512.05369}{{\tt
  arXiv:1512.05369}}].

\bibitem{Schneider:2016uqi}
A.~Schneider, {\it {Astrophysical constraints on resonantly produced sterile
  neutrino dark matter}},  {\em JCAP} {\bf 04} (2016) 059,
  [\href{http://arxiv.org/abs/1601.07553}{{\tt arXiv:1601.07553}}].

\bibitem{Tremaine:1979we}
S.~Tremaine and J.~Gunn, {\it {Dynamical Role of Light Neutral Leptons in
  Cosmology}},  {\em Phys. Rev. Lett.} {\bf 42} (1979) 407--410.

\bibitem{Feng:2018pew}
J.~L. Feng, I.~Galon, F.~Kling, and S.~Trojanowski, {\it {Axionlike particles
  at FASER: The LHC as a photon beam dump}},  {\em Phys. Rev. D} {\bf 98}
  (2018), no.~5 055021, [\href{http://arxiv.org/abs/1806.02348}{{\tt
  arXiv:1806.02348}}].

\bibitem{Dolan:2017osp}
M.~J. Dolan, T.~Ferber, C.~Hearty, F.~Kahlhoefer, and K.~Schmidt-Hoberg, {\it
  {Revised constraints and Belle II sensitivity for visible and invisible
  axion-like particles}},  {\em JHEP} {\bf 12} (2017) 094,
  [\href{http://arxiv.org/abs/1709.00009}{{\tt arXiv:1709.00009}}].

\bibitem{Alekhin:2015byh}
S.~Alekhin et~al., {\it {A facility to Search for Hidden Particles at the CERN
  SPS: the SHiP physics case}},  {\em Rept. Prog. Phys.} {\bf 79} (2016),
  no.~12 124201, [\href{http://arxiv.org/abs/1504.04855}{{\tt
  arXiv:1504.04855}}].

\bibitem{Brdar:2020dpr}
V.~Brdar, B.~Dutta, W.~Jang, D.~Kim, I.~M. Shoemaker, Z.~Tabrizi, A.~Thompson,
  and J.~Yu, {\it {Axion-like Particles at Future Neutrino Experiments: Closing
  the ''Cosmological Triangle''}},  \href{http://arxiv.org/abs/2011.07054}{{\tt
  arXiv:2011.07054}}.

\bibitem{Bauer:2017ris}
M.~Bauer, M.~Neubert, and A.~Thamm, {\it {Collider Probes of Axion-Like
  Particles}},  {\em JHEP} {\bf 12} (2017) 044,
  [\href{http://arxiv.org/abs/1708.00443}{{\tt arXiv:1708.00443}}].

\bibitem{Salvio:2013iaa}
A.~Salvio, A.~Strumia, and W.~Xue, {\it {Thermal axion production}},  {\em
  JCAP} {\bf 01} (2014) 011, [\href{http://arxiv.org/abs/1310.6982}{{\tt
  arXiv:1310.6982}}].

\bibitem{Hall:2009bx}
L.~J. Hall, K.~Jedamzik, J.~March-Russell, and S.~M. West, {\it {Freeze-In
  Production of FIMP Dark Matter}},  {\em JHEP} {\bf 03} (2010) 080,
  [\href{http://arxiv.org/abs/0911.1120}{{\tt arXiv:0911.1120}}].

\bibitem{Heeck:2017xbu}
J.~Heeck and D.~Teresi, {\it {Cold keV dark matter from decays and
  scatterings}},  {\em Phys. Rev. D} {\bf 96} (2017), no.~3 035018,
  [\href{http://arxiv.org/abs/1706.09909}{{\tt arXiv:1706.09909}}].

\bibitem{Braaten:1991dd}
E.~Braaten and T.~C. Yuan, {\it {Calculation of screening in a hot plasma}},
  {\em Phys. Rev. Lett.} {\bf 66} (1991) 2183--2186.

\bibitem{Elahi:2014fsa}
F.~Elahi, C.~Kolda, and J.~Unwin, {\it {UltraViolet Freeze-in}},  {\em JHEP}
  {\bf 03} (2015) 048, [\href{http://arxiv.org/abs/1410.6157}{{\tt
  arXiv:1410.6157}}].

\bibitem{Ballesteros:2020adh}
G.~Ballesteros, M.~A. Garcia, and M.~Pierre, {\it {How warm are non-thermal
  relics? Lyman-$\alpha$ bounds on out-of-equilibrium dark matter}},
  \href{http://arxiv.org/abs/2011.13458}{{\tt arXiv:2011.13458}}.

\bibitem{DEramo:2020gpr}
F.~D'Eramo and A.~Lenoci, {\it {Lower Mass Bounds on FIMPs}},
  \href{http://arxiv.org/abs/2012.01446}{{\tt arXiv:2012.01446}}.

\bibitem{Bolz:2000fu}
M.~Bolz, A.~Brandenburg, and W.~Buchmuller, {\it {Thermal production of
  gravitinos}},  {\em Nucl. Phys. B} {\bf 606} (2001) 518--544,
  [\href{http://arxiv.org/abs/hep-ph/0012052}{{\tt hep-ph/0012052}}]. [Erratum:
  Nucl.Phys.B 790, 336--337 (2008)].

\bibitem{Viel:2005qj}
M.~Viel, J.~Lesgourgues, M.~G. Haehnelt, S.~Matarrese, and A.~Riotto, {\it
  {Constraining warm dark matter candidates including sterile neutrinos and
  light gravitinos with WMAP and the Lyman-alpha forest}},  {\em Phys. Rev. D}
  {\bf 71} (2005) 063534, [\href{http://arxiv.org/abs/astro-ph/0501562}{{\tt
  astro-ph/0501562}}].

\bibitem{Narayanan:2000tp}
V.~K. Narayanan, D.~N. Spergel, R.~Dave, and C.-P. Ma, {\it {Constraints on the
  mass of warm dark matter particles and the shape of the linear power spectrum
  from the Ly$\alpha$ forest}},  {\em Astrophys. J. Lett.} {\bf 543} (2000)
  L103--L106, [\href{http://arxiv.org/abs/astro-ph/0005095}{{\tt
  astro-ph/0005095}}].

\bibitem{Lesgourgues:2011re}
J.~Lesgourgues, {\it {The Cosmic Linear Anisotropy Solving System (CLASS) I:
  Overview}},  \href{http://arxiv.org/abs/1104.2932}{{\tt arXiv:1104.2932}}.

\bibitem{Lesgourgues:2011rh}
J.~Lesgourgues and T.~Tram, {\it {The Cosmic Linear Anisotropy Solving System
  (CLASS) IV: efficient implementation of non-cold relics}},  {\em JCAP} {\bf
  09} (2011) 032, [\href{http://arxiv.org/abs/1104.2935}{{\tt
  arXiv:1104.2935}}].

\bibitem{Aghanim:2018eyx}
{\bf Planck} Collaboration, N.~Aghanim et~al., {\it {Planck 2018 results. VI.
  Cosmological parameters}},  {\em Astron. Astrophys.} {\bf 641} (2020) A6,
  [\href{http://arxiv.org/abs/1807.06209}{{\tt arXiv:1807.06209}}].

\bibitem{Konig:2016dzg}
J.~K\"onig, A.~Merle, and M.~Totzauer, {\it {keV Sterile Neutrino Dark Matter
  from Singlet Scalar Decays: The Most General Case}},  {\em JCAP} {\bf 11}
  (2016) 038, [\href{http://arxiv.org/abs/1609.01289}{{\tt arXiv:1609.01289}}].

\bibitem{Garzilli:2019qki}
A.~Garzilli, O.~Ruchayskiy, A.~Magalich, and A.~Boyarsky, {\it {How warm is too
  warm? Towards robust Lyman-$\alpha$ forest bounds on warm dark matter}},
  \href{http://arxiv.org/abs/1912.09397}{{\tt arXiv:1912.09397}}.

\bibitem{Yeche:2017upn}
C.~Y\`eche, N.~Palanque-Delabrouille, J.~Baur, and H.~du~Mas~des Bourboux, {\it
  {Constraints on neutrino masses from Lyman-alpha forest power spectrum with
  BOSS and XQ-100}},  {\em JCAP} {\bf 06} (2017) 047,
  [\href{http://arxiv.org/abs/1702.03314}{{\tt arXiv:1702.03314}}].

\bibitem{Irsic:2017ixq}
V.~Ir\v{s}i\v{c} et~al., {\it {New Constraints on the free-streaming of warm
  dark matter from intermediate and small scale Lyman-$\alpha$ forest data}},
  {\em Phys. Rev. D} {\bf 96} (2017), no.~2 023522,
  [\href{http://arxiv.org/abs/1702.01764}{{\tt arXiv:1702.01764}}].

\bibitem{Hsueh:2019ynk}
J.-W. Hsueh, W.~Enzi, S.~Vegetti, M.~Auger, C.~D. Fassnacht, G.~Despali, L.~V.
  Koopmans, and J.~P. McKean, {\it {SHARP \textendash{} VII. New constraints on
  the dark matter free-streaming properties and substructure abundance from
  gravitationally lensed quasars}},  {\em Mon. Not. Roy. Astron. Soc.} {\bf
  492} (2020), no.~2 3047--3059, [\href{http://arxiv.org/abs/1905.04182}{{\tt
  arXiv:1905.04182}}].

\bibitem{Bae:2017dpt}
K.~J. Bae, A.~Kamada, S.~P. Liew, and K.~Yanagi, {\it {Light axinos from
  freeze-in: production processes, phase space distributions, and Ly-$\alpha$
  forest constraints}},  {\em JCAP} {\bf 01} (2018) 054,
  [\href{http://arxiv.org/abs/1707.06418}{{\tt arXiv:1707.06418}}].

\bibitem{Murgia:2017lwo}
R.~Murgia, A.~Merle, M.~Viel, M.~Totzauer, and A.~Schneider, {\it {''Non-cold''
  dark matter at small scales: a general approach}},  {\em JCAP} {\bf 11}
  (2017) 046, [\href{http://arxiv.org/abs/1704.07838}{{\tt arXiv:1704.07838}}].

\bibitem{Drlica-Wagner:2019vah}
{\bf DES} Collaboration, A.~Drlica-Wagner et~al., {\it {Milky Way Satellite
  Census. I. The Observational Selection Function for Milky Way Satellites in
  DES Y3 and Pan-STARRS DR1}},  {\em Astrophys. J.} {\bf 893} (2020) 1,
  [\href{http://arxiv.org/abs/1912.03302}{{\tt arXiv:1912.03302}}].

\bibitem{Schneider:2013ria}
A.~Schneider, R.~E. Smith, and D.~Reed, {\it {Halo Mass Function and the Free
  Streaming Scale}},  {\em Mon. Not. Roy. Astron. Soc.} {\bf 433} (2013) 1573,
  [\href{http://arxiv.org/abs/1303.0839}{{\tt arXiv:1303.0839}}].

\bibitem{Schneider:2014rda}
A.~Schneider, {\it {Structure formation with suppressed small-scale
  perturbations}},  {\em Mon. Not. Roy. Astron. Soc.} {\bf 451} (2015), no.~3
  3117--3130, [\href{http://arxiv.org/abs/1412.2133}{{\tt arXiv:1412.2133}}].

\bibitem{Wang:2015ala}
W.~Wang, J.~Han, A.~P. Cooper, S.~Cole, C.~Frenk, and B.~Lowing, {\it
  {Estimating the dark matter halo mass of our Milky Way using dynamical
  tracers}},  {\em Mon. Not. Roy. Astron. Soc.} {\bf 453} (2015), no.~1
  377--400, [\href{http://arxiv.org/abs/1502.03477}{{\tt arXiv:1502.03477}}].

\bibitem{Callingham:2018vcf}
T.~Callingham, M.~Cautun, A.~J. Deason, C.~S. Frenk, W.~Wang, F.~A. G\'omez,
  R.~J. Grand, F.~Marinacci, and R.~Pakmor, {\it {The mass of the Milky Way
  from satellite dynamics}},  \href{http://arxiv.org/abs/1808.10456}{{\tt
  arXiv:1808.10456}}.

\bibitem{Cautun:2019eaf}
M.~Cautun, A.~Benitez-Llambay, A.~J. Deason, C.~S. Frenk, A.~Fattahi, F.~A.
  G\'omez, R.~J. Grand, K.~A. Oman, J.~F. Navarro, and C.~M. Simpson, {\it {The
  Milky Way total mass profile as inferred from Gaia DR2}},  {\em Mon. Not.
  Roy. Astron. Soc.} {\bf 494} (2020), no.~3 4291--4313,
  [\href{http://arxiv.org/abs/1911.04557}{{\tt arXiv:1911.04557}}].

\bibitem{Karukes:2019jwa}
E.~Karukes, M.~Benito, F.~Iocco, R.~Trotta, and A.~Geringer-Sameth, {\it {A
  robust estimate of the Milky Way mass from rotation curve data}},  {\em JCAP}
  {\bf 05} (2020) 033, [\href{http://arxiv.org/abs/1912.04296}{{\tt
  arXiv:1912.04296}}].

\bibitem{Wang:2019ubx}
W.~Wang, J.~Han, M.~Cautun, Z.~Li, and M.~N. Ishigaki, {\it {The mass of our
  Milky Way}},  {\em Sci. China Phys. Mech. Astron.} {\bf 63} (2020), no.~10
  109801, [\href{http://arxiv.org/abs/1912.02599}{{\tt arXiv:1912.02599}}].

\bibitem{Dvali:1995ce}
G.~Dvali, {\it {Removing the cosmological bound on the axion scale}},
  \href{http://arxiv.org/abs/hep-ph/9505253}{{\tt hep-ph/9505253}}.

\bibitem{Co:2018phi}
R.~T. Co, E.~Gonzalez, and K.~Harigaya, {\it {Axion Misalignment Driven to the
  Bottom}},  {\em JHEP} {\bf 05} (2019) 162,
  [\href{http://arxiv.org/abs/1812.11186}{{\tt arXiv:1812.11186}}].

\bibitem{Adhikari:2016bei}
M.~Drewes et~al., {\it {A White Paper on keV Sterile Neutrino Dark Matter}},
  {\em JCAP} {\bf 01} (2017) 025, [\href{http://arxiv.org/abs/1602.04816}{{\tt
  arXiv:1602.04816}}].

\bibitem{Planck:2018jri}
{\bf Planck} Collaboration, Y.~Akrami et~al., {\it {Planck 2018 results. X.
  Constraints on inflation}},  {\em Astron. Astrophys.} {\bf 641} (2020) A10,
  [\href{http://arxiv.org/abs/1807.06211}{{\tt arXiv:1807.06211}}].

\bibitem{Boyarsky:2007ge}
A.~Boyarsky, D.~Malyshev, A.~Neronov, and O.~Ruchayskiy, {\it {Constraining DM
  properties with SPI}},  {\em Mon. Not. Roy. Astron. Soc.} {\bf 387} (2008)
  1345, [\href{http://arxiv.org/abs/0710.4922}{{\tt arXiv:0710.4922}}].

\bibitem{Roach:2019ctw}
B.~M. Roach, K.~C. Ng, K.~Perez, J.~F. Beacom, S.~Horiuchi, R.~Krivonos, and
  D.~R. Wik, {\it {NuSTAR Tests of Sterile-Neutrino Dark Matter: New Galactic
  Bulge Observations and Combined Impact}},  {\em Phys. Rev. D} {\bf 101}
  (2020), no.~10 103011, [\href{http://arxiv.org/abs/1908.09037}{{\tt
  arXiv:1908.09037}}].

\bibitem{Horiuchi:2013noa}
S.~Horiuchi, P.~J. Humphrey, J.~Onorbe, K.~N. Abazajian, M.~Kaplinghat, and
  S.~Garrison-Kimmel, {\it {Sterile neutrino dark matter bounds from galaxies
  of the Local Group}},  {\em Phys. Rev. D} {\bf 89} (2014), no.~2 025017,
  [\href{http://arxiv.org/abs/1311.0282}{{\tt arXiv:1311.0282}}].

\bibitem{Nadler:2020prv}
{\bf DES} Collaboration, E.~Nadler et~al., {\it {Milky Way Satellite Census.
  III. Constraints on Dark Matter Properties from Observations of Milky Way
  Satellite Galaxies}},  \href{http://arxiv.org/abs/2008.00022}{{\tt
  arXiv:2008.00022}}.

\bibitem{Viaux:2013lha}
N.~Viaux, M.~Catelan, P.~B. Stetson, G.~Raffelt, J.~Redondo, A.~A.~R. Valcarce,
  and A.~Weiss, {\it {Neutrino and axion bounds from the globular cluster M5
  (NGC 5904)}},  {\em Phys. Rev. Lett.} {\bf 111} (2013) 231301,
  [\href{http://arxiv.org/abs/1311.1669}{{\tt arXiv:1311.1669}}].

\bibitem{Bertolami:2014wua}
M.~M. Miller~Bertolami, B.~E. Melendez, L.~G. Althaus, and J.~Isern, {\it
  {Revisiting the axion bounds from the Galactic white dwarf luminosity
  function}},  {\em JCAP} {\bf 10} (2014) 069,
  [\href{http://arxiv.org/abs/1406.7712}{{\tt arXiv:1406.7712}}].

\bibitem{Craig:2018kne}
N.~Craig, A.~Hook, and S.~Kasko, {\it {The Photophobic ALP}},  {\em JHEP} {\bf
  09} (2018) 028, [\href{http://arxiv.org/abs/1805.06538}{{\tt
  arXiv:1805.06538}}].

\bibitem{Benso:2019jog}
C.~Benso, V.~Brdar, M.~Lindner, and W.~Rodejohann, {\it {Prospects for Finding
  Sterile Neutrino Dark Matter at KATRIN}},  {\em Phys. Rev. D} {\bf 100}
  (2019), no.~11 115035, [\href{http://arxiv.org/abs/1911.00328}{{\tt
  arXiv:1911.00328}}].

\bibitem{Drlica-Wagner:2019xan}
{\bf LSST Dark Matter Group} Collaboration, A.~Drlica-Wagner et~al., {\it
  {Probing the Fundamental Nature of Dark Matter with the Large Synoptic Survey
  Telescope}},  \href{http://arxiv.org/abs/1902.01055}{{\tt arXiv:1902.01055}}.

\bibitem{Dvorkin:2020xga}
C.~Dvorkin, T.~Lin, and K.~Schutz, {\it {The cosmology of sub-MeV dark matter
  freeze-in}},  \href{http://arxiv.org/abs/2011.08186}{{\tt arXiv:2011.08186}}.

\end{thebibliography}\endgroup

\end{document}